\documentclass[preprint,showpacs,preprintnumbers,superscriptaddress]{revtex4-1}
\usepackage{multirow}
\usepackage{diagbox}
\usepackage{booktabs}
\usepackage{array}
\usepackage{natbib}
\usepackage{amsmath,amssymb,graphicx,bm,subfigure,float,siunitx,color}
\usepackage{bm,subfigure,float,siunitx,graphicx}
\usepackage[figuresright]{rotating}
\usepackage{color}
\usepackage{epstopdf}
\usepackage{mathrsfs}
\newcommand{\be}{\begin{equation}}
\newcommand{\ee}{\end{equation}}

\newcommand{\1}{\left}
\newcommand{\2}{\right}

\begin{document}
\title{\boldmath Observational signatures of Bardeen black holes under nonlinear electrodynamics }

\author{Haiyuan Feng\footnote{Corresponding author}}
\email{Email address:  fenghaiyuan@sxnu.edu.cn}
\affiliation{School of Physics and Electronic Engineering, Shanxi Normal University, Taiyuan 030031, China}

\author{Dan-Dan Cui\footnote{Corresponding author}}
\email{Email address: cuidandan0728@163.com  }
\affiliation{School of Physics and Electronic Engineering, Shanxi Normal University, Taiyuan 030031, China}

\begin{abstract}
We investigate the shadow and optical appearance of the Bardeen black hole (BH) within nonlinear electrodynamics (NLED). Since photon propagation in  nonlinear electromagnetic field can be equivalently described as null-geodesic motion in the effective metric induced by NLED, we derive the corresponding effective optical geometry of the NLED-Bardeen BH. Furthermore, by examining the horizon–photon sphere–shadow inequality, we demonstrate that the NLED-Bardeen BH does not satisfy the complete inequality chain. 
In addition, we investigate photon dynamics and the corresponding optical images generated by both thin disk and spherical accretion models. Increasing \(g/M\) reduces the photon sphere radius and critical impact parameter, resulting in a more compact shadow and an inward shift of the emission peak. For the thin disk accretion model, photon trajectories are classified into direct emission, lensing ring, and photon ring components according to the corresponding impact parameter. We find that Model 1 exhibits three distinct intensity peaks, whereas Models 2 and 3 display two and one peaks. The systematic dependence of these features on the magnetic charge indicates that the NLED-Bardeen BH may exhibit observational signatures distinguishable from those of the Bardeen BH. In static spherical accretion, the intensity peaks near the critical impact parameter because of the enhanced optical path length, while infalling spherical accretion introduces an additional Doppler redshift that suppresses the observed intensity. These results indicate that the shadow size, ring structure, and intensity distribution can serve as potential observational signatures of NLED effects in the strong-field regime.

\end{abstract}

\maketitle
\section{Introduction}
The direct imaging of the supermassive black holes (BHs) M87* and Sgr A* by the Event Horizon Telescope (EHT) has opened a new observational window for studying gravity and compact objects in the strong-field regime \cite{EventHorizonTelescope:2019dse,EventHorizonTelescope:2019uob,EventHorizonTelescope:2019jan,EventHorizonTelescope:2019ths,EventHorizonTelescope:2019pgp,EventHorizonTelescope:2019ggy,EventHorizonTelescope:2022wkp,EventHorizonTelescope:2022apq,EventHorizonTelescope:2022wok,EventHorizonTelescope:2022exc,EventHorizonTelescope:2022urf,EventHorizonTelescope:2022xqj}. 
A central theoretical object in BH imaging is the shadow, which is generated by the unstable photon region and by strong gravitational lensing near the compact object \cite{Takahashi:2005hy,Luminet:1979nyg,Lu:2014zja,Gralla:2019xty,Narayan:2019imo}. 
Geometrically, the shadow boundary can be identified with the critical curve on the observer's screen, separating photon trajectories that escape to infinity from those captured by the BH \cite{Dokuchaev:2019jqq}. Since this boundary is determined by near-horizon photon dynamics, the shadow morphology, together with the associated lensing-ring and photon-ring structures, encodes key properties of the compact object, including its mass, angular momentum, and possible deviations from the Kerr geometry \cite{Broderick:2008qf,Hioki:2009na,Bambi:2019tjh,Gralla:2020srx,Tsukamoto:2014tja}. 
BH shadows and related image features have therefore been widely used to investigate accretion disk physics \cite{Cunha:2019hzj,Abramowicz:2011xu,Zhang:2024lsf,Feng:2024iqj}, dark matter effects \cite{Haroon:2018ryd,Konoplya:2019sns,Chen:2019fsq,Faraji:2025efg}, cosmological backgrounds \cite{Li:2020drn,Chowdhuri:2020ipb}, and modified gravity or non-Kerr compact objects \cite{Wei:2013kza,Feng:2025ljz,Cai:2025amk,Cai:2025pan,Wu:2025hcu,Cunha:2015yba,Cunha:2016wzk,Wei:2020ght,Vagnozzi:2019apd}. 
In these studies, the observed image is determined not only by the spacetime metric, but also by photon propagation and by the structure of the emitting matter around the BH. 
This makes shadows, lensing rings, photon rings, and disk intensity profiles sensitive diagnostics of the underlying gravitational and electromagnetic environment.

The morphology of the BH shadows is fundamentally dictated by the null trajectories of photons within the underlying spacetime. In an idealized vacuum environment, under the approximation of geometric optics, photons follow null geodesics, such that their equations of motion are governed strictly by the spacetime metric. However, in realistic astrophysical settings, the presence of fields beyond the purely gravitational sector is inevitable. A more plausible physical scenario considers the interaction between photons—acting as electromagnetic entities-and auxiliary fields during their traversal from the BH vicinity to a distant observer. In particular, magnetic fields play a pivotal role in the magnetohydrodynamics of BH environments \cite{Abramowicz:2011xu,Eatough:2013nva,Doeleman:2012zc}, and mounting observational evidence—most notably from polarized EHT data—confirms the existence of robust, large-scale magnetic fields surrounding supermassive BHs.  From the theoretical side, magnetic fields and nonlinear electromagnetic interactions can affect photon propagation in different ways. For instance, BH shadows in external magnetic universes have been investigated for Schwarzschild and Kerr backgrounds \cite{Junior:2021dyw,Wang:2021ara}. 
These models are useful for isolating magnetic-field effects; however, the gravitational backreaction of the external field generally changes the global asymptotic structure of the spacetime. Another possibility, which is the focus of the present work, is that nonlinear electrodynamics (NLED) modifies the optical geometry. In NLED, electromagnetic disturbances generally propagate along null directions of an effective metric determined by the background electromagnetic field and by the derivatives of the NLED Lagrangian \cite{DeLorenci:2000yh,Novello:1999pg,Hu:2020usx,Zhong:2021mty}. Thus, the photon trajectories relevant for BH imaging differ from the null geodesics of the background spacetime metric, leading to potentially observable changes in the shadow size, lensing structure, and image morphology. While this approach provides a tractable framework for isolating the influence of magnetic fields on shadow morphology, it introduces a significant theoretical caveat: the resulting spacetime manifold loses its property of asymptotic flatness, a departure from the standard isolated BH solutions.

Additionally, in General Relativity (GR), the paradigm of continuous gravitational collapse inevitably leads to the existence of singularities—regions of divergent curvature where the classical description of spacetime breaks down. Addressing this foundational challenge has motivated a substantial body of research dedicated to the construction of regular BH solutions. This trajectory began with the seminal work of Bardeen \cite{1968qtr..conf...87B}, who in 1968 proposed a BH geometry that remains non-singular at the origin. Physically, this model can be interpreted as gravitationally collapsed magnetic monopole coupled with a specific class of nonlinear electrodynamics (NLED) \cite{Ayon-Beato:2000mjt}. Following this pioneering approach, researchers have sought to regularize the central singularity by either modifying the gravitational action or introducing exotic matter contents. These efforts have yielded several well-known non-singular models, most notably the Ayón-Beato and García BH \cite{Ayon-Beato:1998hmi}, the Hayward BH \cite{Hayward:2005gi}, and the Berej-Matyjasek–Trynieki-Wornowicz BH \cite{Berej:2006cc}. These spacetimes provide useful theoretical laboratories for investigating how nonlinear matter sources modify strong-field observables.

Although the Bardeen geometry itself originates from a nonlinear electromagnetic source, many shadow studies treat photons as test particles following null geodesics of the background Bardeen metric. However, if the electromagnetic field is governed by NLED, photon propagation should be described by the corresponding effective optical geometry. 
This distinction leads to two conceptually different descriptions: the Bardeen BH, in which photons follow the null geodesics of the background metric, and the NLED-Bardeen BH, in which photons follow the null geodesics of the NLED induced effective metric. Comparing these two descriptions is important for identifying which observational features are caused by the regular Bardeen geometry itself and which are genuine consequences of nonlinear electromagnetic corrections to photon propagation.

The structure of this paper is outlined as follows. In Section II, we introduce the Bardeen BH within the framework of NLED and analyze its spacetime geometry, including the conditions for the existence of event horizons. In Section III, we consider the NLED effects to provide an NLED-Bardeen BH solution, which serves as the foundational basis for our subsequent research. Section IV is devoted to the propagation of photons in the NLED-corrected effective optical geometry. We analyze the photon sphere and shadow radii and examine whether the NLED-Bardeen BH satisfies the horizon–photon sphere–shadow inequalities. 
We also investigate how the magnetic charge $g$ modifies the photon trajectories and quantitatively compare the associated optical observables. In Section V, we study the observational signatures of a geometrically thin accretion disk, with particular emphasis on the corresponding two-dimensional images. Section VI is devoted to the optical appearance of the NLED-Bardeen BH surrounded by both static and radially infalling spherical accretion flows. Furthermore, we comparatively analyzed the observed intensity and optical appearance of the three BH configurations, together with the influence of the magnetic charge on the intensity distribution.
Finally, the last section summarizes our principal results and presents the main conclusions of this work.

\section{Spherical symmetric Bardeen black hole }

The Bardeen BH serves as a quintessential paradigm of a regular, singularity-free spacetime. Originally formulated by Bardeen in 1968 \cite{1968qtr..conf...87B}, this metric was introduced as a conceptual counterexample to the prevailing view—derived from the singularity theorems—that gravitational collapse necessarily culminates in a curvature singularity shielded by an event horizon. While initially regarded as a phenomenological construct, its physical viability was significantly bolstered decades later by Ayón-Beato and García \cite{Ayon-Beato:2000mjt}. They demonstrated that the Bardeen geometry arises as an exact solution to the Einstein equations coupled with NLED, representing the gravitational field of a self-gravitating magnetic monopole. 
From physical perspective, which is described by the Lagrangian \( \mathcal{L}(F) \)
\be
\label{1}
\mathcal{L}(F)=\frac{3 M}{|g|^3}\left(\frac{\sqrt{g^2 F/2}}{1+\sqrt{ g^2 F/2}}\right)^{\frac{5}{2}},
\ee
which satisfies \( F_{\mu\nu} = \nabla_{\mu}A_{\nu} - \nabla_{\nu}A_{\mu} \), with \( \mathcal{L}(F) \) being a function of the electromagnetic invariant \( F_{\mu\nu}F^{\mu\nu} \). Furthermore, the electromagnetic configuration is characterized by the four-vector potential $A_{\mu} = g\cos\theta \delta^{\phi}_{\mu}$, where the parameter $g$ signifies the magnetic charge that sources the nonlinear field \cite{Ayon-Beato:2000mjt,Zhang:2020mxi}. This reinterpretation effectively bridged the gap between mathematical curiosity and theoretical consistency, establishing the Bardeen metric as a robust spacetime endowed with a well-defined, albeit nonlinear, electromagnetic source. Consequently, it has become a fundamental framework for investigating how non-singular cores modify the observational signatures of BHs, such as their shadows and lensing profiles.

The spherical symmetric Bardeen BH characterized by the following metric
\be
\label{2}
ds^2 = -\left(1 - \frac{2Mr^2}{(r^2 + g^2)^{3/2}}\right) dt^2+\left(1 - \frac{2Mr^2}{(r^2 + g^2)^{3/2}}\right)^{-1} dr^2 + r^2 d\Omega^2,
\ee
here the parameters $M$ correspond to the ADM mass. The angular sector of the spacetime is described by the  $d\Omega^2 = d\theta^2 + \sin^2\theta d\phi^2$, which corresponds to the standard line element on a unit two-sphere. This term encapsulates the fundamental spherical symmetry of the background geometry, ensuring that the metric remains invariant under rotations within the $\rm SO(3)$ isometry group. It is noteworthy that the Bardeen metric exhibits distinct limiting behaviors depending on the radial scale. In the far-field limit ($g/r \ll 1$), the spacetime is asymptotically flat and converges to the Schwarzschild geometry, as the influence of the nonlinear electromagnetic charge vanishes. Conversely, in the high-curvature regime near the coordinate origin ($g/r \gg 1$), the metric transitions into a de Sitter-like geometry.
\be
\label{3}
ds^2 = -\left(1 - \frac{2M}{g^{3}}r^2\right) dt^2+\left(1 - \frac{2M}{g^{3}}r^2\right)^{-1} dr^2 + r^2 d\Omega^2,
\ee

As evidenced by Eq. \eqref{3}, the Bardeen BH circumvents the formation of spacetime singularity through the emergence of de Sitter-like core. This interior structure is characterized by negative pressure, which provides the necessary repulsive force to counteract gravitational collapse and preclude a singular final state \cite{Sakharov:1966hcb,Nam:2018uvc}. A significant feature of such regular BH is their inherent violation of the weak energy condition (WEC). However, this non-classical behavior is strictly confined to the deep interior, as the region of violation remains perpetually sequestered within the domain bounded by the Cauchy horizon \cite{Neves:2014aba,Kumar:2019pjp}. This departure from classical energy conditions is generally interpreted as a phenomenological manifestation of underlying quantum gravity corrections, which act to regularize the high-curvature regime of the BH interior \cite{Kumar:2019pjp}.

Additionally, in contrast to the Reissner–Nordström BH, where the electromagnetic sector contributes to the gravitational potential at order $\mathcal{O}(r^{-2})$, the Bardeen modification introduces a more rapid $\mathcal{O}(r^{-3})$ decay. This higher-order suppression of the magnetic charge's influence at the core is the fundamental mechanism that regularizes the BH; it ensures the regularity of all essential curvature invariants—specifically the Ricci scalar $R$, the Ricci tensor square $R_{\mu\nu}R^{\mu\nu}$, and the Kretschmann scalar $R_{\mu\nu\alpha\beta}R^{\mu\nu\alpha\beta}$—which remain nonsingular as $r \to 0$. Consequently, the Bardeen metric successfully resolves the coordinate singularity inherent in classical solutions while preserving the causal structure necessary for event horizons. In the analysis of the horizon structure, the Bardeen BH exhibits a more intricate behavior determined by the ratio of the magnetic charge $g$ to the ADM mass $M$. A systematic analysis of the metric function—specifically by simultaneously solving $f(r, g) = 0$ and $f'(r, g) = 0$—reveals a critical threshold at $g/M \approx 0.7698$ \cite{Cui:2025qxu}. This value represents the extremal limit where the Cauchy (inner) horizon $r_-$ and the event (outer) horizon $r_+$ coalesce at a degenerate radius of $r_\ast/M \approx 1.08866$. Consequently, for sub-critical configurations where $g/M \in (0, 0.7698)$, the spacetime admits two distinct horizons. However, should the charge-to-mass ratio surpass this critical bound, the horizons vanish, transitioning the geometry into a horizonless, non-singular compact object.

\section{Effective geometry induced by nonlinear electrodynamics}
In this section, we employ the formal framework established by Novello et al. \cite{Novello:1999pg,DeLorenci:2000yh,Bronnikov:2000vy} to derive the effective geometry induced by NLED effects. This approach accounts for the fact that NLED interactions effectively modify the background spacetime perceived by photons, causing them to propagate along modified trajectories rather than the standard null geodesics of the underlying metric. Guided by the Bardeen-type Lagrangian introduced in Eq. \eqref{1}, we consider a general NLED Lagrangian of the form $L=L(F)$, which is a function of the electromagnetic invariants 
$F_{\mu\nu}F^{\mu\nu}$. By applying the variational principle to this Lagrangian density, the resulting field equations of motion are expressed as
\be
\label{4}
\partial_{\mu}(L_FF^{\mu\nu} )=0,
\ee
where we restrict our analysis to a flat spacetime background, noting that the resulting conclusions can be naturally extended to BH via the principle of general covariance. We consider a surface of constant phase, denoted by $\Sigma$, which serves as a boundary of discontinuity for the electromagnetic field derivatives. Thus, we require the electromagnetic field strengths $F_{\alpha\beta}$ to remain continuous across $\Sigma$, while their first-order derivatives exhibit a jump discontinuity. By imposing the formal matching conditions $[F_{\alpha\beta}]_\Sigma = 0$ and $[\partial_\sigma F_{\alpha\beta}]_\Sigma = f_{\alpha\beta}k_\sigma$ upon the surface.
$[F_{\alpha\beta}]_\Sigma \equiv F_{\alpha\beta}^+ - F_{\alpha\beta}^-$ denotes the discrete jump across the discontinuity surface $\Sigma$, where the tensor $f_{\alpha\beta}$ characterizes the discontinuities of the field on the surface $\Sigma$ and $k_\lambda$ signifies the wave propagation four-vector. Furthermore, to satisfy the cyclic constraints imposed by the Bianchi identity, these quantities must obey the relation
\be
\label{5}
f_{\mu\nu}k_{\lambda} + f_{\nu\lambda}k_{\mu} +  f_{\lambda\mu}k_{\nu} = 0.
\ee

In addition, the field equations of motion presented in Eq. \eqref{4} reduce to
\be
\label{6}
L_F f^{\mu\nu}k_{\nu}+2L_{FF}a F^{\mu\nu}k_{\nu}=0,
\ee
where $a$ is defined as
\be
\label{7}
a\equiv F^{\alpha\beta}f_{\alpha \beta}.
\ee

To derive a scalar identity, we contract the aforementioned expression with $k_{\alpha}\eta^{\alpha \lambda}F^{\mu\nu}$, which yields
\be
\label{8}
a k_\nu k_\mu \eta^{\mu \nu}+2 F^{\mu \nu} f_\nu{ }^\lambda k_\lambda k_\mu=0.
\ee
By combining Eqs. \eqref{6} and \eqref{8}, we derive the propagation equation governing the field discontinuities, which is expressed as \cite{DeLorenci:2000yh}
\be
\label{9}
(g^{\mu\nu}_{\rm eff})k_{\mu}k_{\nu}=(L_F\eta^{\mu\nu}-4L_{FF}F^{\mu\alpha}F_{\alpha}^{\nu})k_{\mu}k_{\nu}=0.
\ee

Equation \eqref{9} implies that the self-interaction of the background field $F_{\mu\nu}$ fundamentally alters the propagation dynamics of electromagnetic discontinuities. This interaction can be phenomenologically interpreted as an induced modification of the underlying metric $\eta_{\mu\nu}$, effectively giving rise to a new, emergent geometry. A closer examination of this relation reveals that the standard propagation of field discontinuities along the null geodesics of the flat background occurs exclusively within the regime of linear Maxwell electrodynamics; any departure from linearity shifts the trajectories to the null paths of the effective spacetime. Therefore, for the Bardeen BH described by Eq. \eqref{2} which originates from the nonlinear electromagnetic Lagrangian in Eq. \eqref{1}, it is naturally essential to introduce an effective metric to accurately characterize the photon trajectories. This effective framework serves as the requisite background for deriving the equations of motion under the influence of NLED corrections.

To explicitly manifest the functional dependence of the effective metric on the energy-momentum tensor, we consider Lagrangian density $\mathcal{L}(F)$, for which the following relations hold
\be
\label{10}
T_{\mu\nu}=-4\mathcal{L}_FF_{\mu}^{\alpha}F_{\alpha\nu}-\mathcal{L}\eta_{\mu\nu}.
\ee

In terms of this tensor the effective geometry \eqref{9} can be expanded as
\be
\label{11}
g^{\mu\nu}_{\rm eff}=\left(\mathcal{L}_F+\frac{\mathcal{L}\mathcal{L}_{FF}}{\mathcal{L}_{F}}\right)\eta^{\mu\nu}+\frac{\mathcal{L}_{FF}}{\mathcal{L}_F}T^{\mu\nu}.
\ee

We emphasize that since the modified trajectories of photons are governed by the energy-momentum tensor of the background field, the resulting spacetime manifold perceived by the radiation is inherently coupled to the nonlinear electromagnetic source. Consequently, field discontinuities propagate along the null geodesics of an effective geometry that is functionally dependent on the local field energy distribution. It is important to underscore, as evidenced by the preceding derivation, that the energy distribution of the electromagnetic field acts as the fundamental source of geometric deformation. This coupling ensures that, from the perspective of propagating photons, the effective manifold deviates from the standard Minkowskian background; this deviation is a direct manifestation of the nonlinear self-interactions inherent in the NLED framework, where the field effectively acts as its own medium of refraction.

Since the effective BH is formally defined via the inverse tensor $g_{\rm eff}^{\mu\nu}$, we utilize the fundamental orthogonality relation, $g_{\rm eff}^{\mu\rho} g_{\nu\rho}^{\rm eff} = \delta^{\mu}_{\nu}$, to determine the covariant components of the effective metric. Applying this procedure to the Bardeen BH described by Eq. \eqref{2}, we derive the explicit form of the effective BH as perceived by photon propagation. Specifically, the non-zero components of the effective metric are obtained as
\be
\label{12}
ds^2_{\rm eff} = \frac{1}{\mathcal{L}_F}\left(-\left(1 - \frac{2Mr^2}{(r^2 + g^2)^{3/2}}\right) dt^2+\left(1 - \frac{2Mr^2}{(r^2 + g^2)^{3/2}}\right)^{-1} dr^2\right) + \frac{r^2}{\mathcal{L}_F+2F \mathcal{L}_{FF}} d\Omega^2,
\ee
here $\mathcal{L}_F$ denotes the partial derivative of the Lagrangian density $\mathcal{L}(F)$ with respect to the invariant $F$, as defined in Eq. \eqref{1}. The contribution of the electromagnetic corrections within the denominator can be explicitly evaluated as

\be
\1\{\begin{split}
\label{13}
&\mathcal{L}_F=\frac{15 M r^6 \sqrt{r^2+g^2}}{8 \left(g^2+r^2\right)^4},\\
&\mathcal{L}_F+2F \mathcal{L}_{FF}=\frac{15 M r^6 \left(3 r^2-4 g^2\right) \sqrt{r^2+g^2}}{16 \left(g^2+r^2\right)^5},
\end{split}\2.
\ee
where we employ the electromagnetic invariant $F = 2g^2/r^4$ as the foundational basis. Consequently, we have derived the analytical expressions for the effective metric incorporating NLED effects. This refined spacetime framework accounts for the self-interaction of the electromagnetic field, which fundamentally deviates from the linear Maxwellian regime. 
From Eq. \eqref{12}, several important physical consequences of the NLED correction can be identified.

\begin{figure}[htph]
	\begin{minipage}{1.0\textwidth}
    \centering
		\includegraphics[scale=1,angle=0]{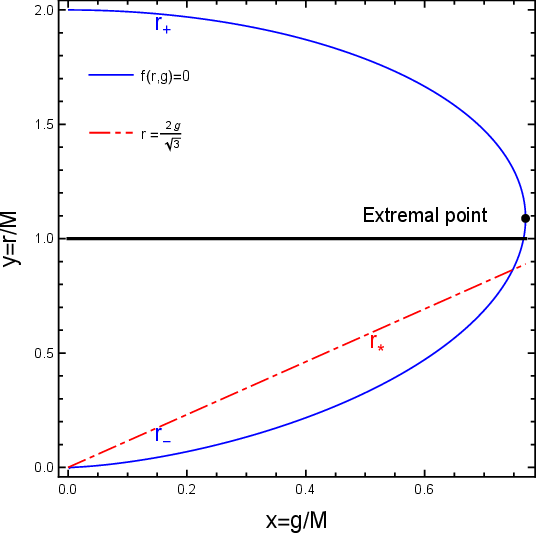}
	\end{minipage}
    \hspace{4cm}
\caption{Figure shows the parameter space configuration of the inner and outer horizons together with the critical radius in the NLED-Bardeen BH. The blue curves represent the inner and outer horizons, the red dashed curve corresponds to the critical radius $r_{\star}$, and the black solid point denotes the extremal BH configuration.
}\label{Fig.1}
\end{figure}

\begin{itemize}

\item The NLED-induced Bardeen BH (NLED-Bardeen BH) does not modify the background Bardeen geometry itself. In particular, the  function, $f(r)\equiv1-2Mr^{2}/(r^{2}+g^{2})^{3/2}$, remains unchanged. Consequently, the location of the outer event horizon, determined by $f(r_+)=0$, coincides with that of the original Bardeen BH, provided that $\mathcal{L}_{F}$ is finite and nonvanishing at the horizon. This result is physically expected, since the effective metric describes the propagation of electromagnetic perturbations in the nonlinear medium rather than a backreaction-induced modification of the underlying spacetime geometry. Therefore, the intrinsic horizon structure of the Bardeen solution are preserved.

\item The allowed range of the magnetic charge parameter $g$ is still governed by the horizon condition of the background Bardeen BH. In particular, the extremal configuration is determined by $g/M\leq 0.7698$, and the NLED optical correction introduces no additional modification to this upper bound. Hence, the parameter $g$ continues to characterize the magnetic charge and the regularization scale of the background geometry, while the NLED effect primarily changes the optical properties of photon propagation.

\item The photon ring radius, shadow size, strong lensing signatures, and higher-order images are all expected to be substantially modified because NLED changes the effective light-cone structure governing photon propagation. Consequently, all observables associated with the optical properties of the spacetime can be affected. These distinct optical signatures may therefore provide a potential observational means of discriminating between the Bardeen BH and the NLED-Bardeen BH.

\item Eq. \eqref{13} shows that $\mathcal{L}_{F}+2F\mathcal{L}_{FF}\propto 3r^{2}-4g^{2}$, which vanishes at the critical radius $r_{\star}=\frac{2|g|}{\sqrt{3}}$. At this radius, the angular component of the NLED-Bardeen BH becomes singular, while for $r<r_{\star}$ its sign changes. This indicates that the conventional static optical interpretation of the effective geometry breaks down: the angular directions cannot be interpreted as ordinary spacelike directions for electromagnetic perturbations. Equivalently, the characteristic structure of photon propagation becomes degenerate at $r=r_{\star}$. Therefore, the physically reliable optical analysis should be restricted to the region $r>r_{\star}$, where the effective metric retains the standard causal interpretation relevant for photon motion. Importantly, Fig. \ref{Fig.1} presents the inner and outer horizons together with the critical radius $r_{\star}$. It clearly shows that, throughout the Bardeen BH branch, the critical radius always lies inside the outer horizon, namely $r_{\star}<r_{+}$, including in the extremal limit. Consequently, the exterior photon region relevant to BH shadow formation and gravitational lensing remains well defined.
\end{itemize}

\section{Null geodesic equation and shadow cast}
The photon sphere, defined by the locus of unstable circular null geodesics, serves as the fundamental separator between the capture zone of the BH and the asymptotic region. Consequently, the apparent shadow perceived by a distant observer is a direct projection of these critical trajectories, providing a visual signature of the underlying spacetime geometry. Within the framework of the NLED-Bardeen spacetime—as characterized by the modified metric in Eq. \eqref{12}—the non-trivial self-interaction of the electromagnetic field introduces significant corrections to the optical properties of the Bardeen BH. Specifically, the nonlinearity of the Lagrangian fundamentally alter the radius of the photon sphere and modulating the resulting shadow size.

To investigate the dynamical behavior of photons within this framework, we express the effective metric from Eq. \eqref{12} in a form conducive to the analysis of the equations of motion \cite{Uniyal:2022vdu,Perlick:2015vta}. Specifically, the line element of the NLED-Bardeen BH can be written as
\be
\label{14}
ds^2_{\rm eff} =\frac{1}{\mathcal{L}_F}\left( -A(r) dt^2+B(r) dr^2 + C(r) d\Omega^2 \right),
\ee
with
\be
\1\{\begin{split}
\label{15}
&A(r)\equiv f(r)=1 - \frac{2Mr^2}{(r^2 + g^2)^{3/2}}, \\
&B(r)\equiv f^{-1}(r),\\
&C(r)\equiv  \frac{r^2\mathcal{L}_F}{\mathcal{L}_F+2F \mathcal{L}_{FF}}.
\end{split}\2.
\ee

Since photons propagate along null geodesics, their trajectories are invariant under conformal transformations of the metric. Therefore, the overall conformal factor $1/\mathcal{L}_{F}$ in the effective metric does not affect the null paths, the photon sphere condition, or the associated critical impact parameter determined from the optical geometry. In the following analysis, we therefore omit this common conformal factor and work with the conformally related effective metric, which preserves the relevant null geodesic structure while considerably simplifying the calculation.
By exploiting the spherical symmetry of the effective metric \eqref{14}, we may restrict our analysis to the equatorial plane without loss of generality. By fixing the polar angle at $\theta = \pi/2$ (and consequently $p_\theta = 0$), the metric components are restricted to the equatorial plane. Within this reduced dimensionality, the dynamics of light rays are governed by the Hamiltonian for null geodesics, which is defined as
\be
\label{16}
H=\frac{1}{2}g^{\mu\nu}_{\rm eff}p_{\mu}p_{\nu}=\frac{1}{2}\left( -\frac{p^2_t}{A(r)}+\frac{p^2_r}{B(r)}+\frac{p^2_\phi}{C(r)} \right),
\ee
where $p_{t}$, $p_{r}$, and $p_{\phi}$ denote the components of the photon's four-momentum, which are conjugate to the coordinates $(t, r, \phi)$. The dynamic evolution of these variables is governed by the Hamilton's canonical equations, expressed as
\begin{equation}
 \left\{\begin{split}
&\dot{x}^{\mu}=\frac{\partial H}{\partial p_{\mu}},\\
&\dot{p}_{\mu}=-\frac{\partial H}{\partial x^{\mu}}.\\
	\end{split}\right. \label{17}  
\end{equation}
here, $\dot{x} \equiv dx/d\lambda$ denotes the derivative with respect to the affine parameter $\lambda$. 
Furthermore, we can derive
\begin{equation}
 \left\{\begin{split}
&\dot{t}=-\frac{p_t}{A(r)}, \quad \dot{\phi}=\frac{p_\phi}{C(r)}, \quad \dot{r}=-\frac{p_r}{B(r)}, \\
&\dot{p}_t=0,\quad\dot{p}_\phi=0,\\
&\dot{p}_r=\frac{1}{2}\left(-\frac{p_t^2 A^{\prime}(r)}{A(r)^2}+\frac{p_r^2 B^{\prime}(r)}{B(r)^{2}}-\frac{p^2_{\phi}C^{\prime}(r)}{C(r)^2}\right),\\
	\end{split}\right. \label{18}
\end{equation}
where the six first-order differential equations admit two conserved quantities. We define $p_t = -\omega_0$, where $\omega_0$ denotes the conserved energy of the photon along the geodesic. To analyze the photon trajectory confined to a plane, we obtain
\begin{equation}
\frac{dr}{d\phi}= \frac{\dot{r}}{\dot{\phi}}=\pm\sqrt{\frac{C(r)}{B(r)}}  \sqrt{\frac{\omega^2_{0}}{p^2_{\phi}}h(r)^2-1}, \label{19}
\end{equation}
with $h(r)^2\equiv\frac{C(r)}{A(r)}=\frac{\mathcal{L}_F}{\mathcal{L}_F+2F\mathcal{L}_{FF}}\left(\frac{r^2}{1-2Mr^2/(r^2+g^2)^{3/2}}\right)$. It should be noted that the null constraint condition, $H = 0$, has been explicitly imposed throughout the derivation, consistent with the propagation of massless particles. For a stationary circular orbit at a constant radius, the radial velocity and acceleration must vanish simultaneously, i.e., $\dot{r} = 0$ and $\ddot{r} = 0$, which consequently implies $p_r = 0$. By imposing these equilibrium conditions, Eq. \eqref{18} reduces to the following constraints for the critical orbits
\begin{equation}
 \left\{\begin{split}
&0=-\frac{\omega^2_{0}}{A(r)}+\frac{p^2_{\phi}}{C(r)},\\
&0=\frac{\omega^2_{0} A^{\prime}(r)}{A(r)^2}+\frac{p^2_{\phi}C^{\prime}(r)}{C(r)^2}.
\end{split}\right. \label{20}
\end{equation}

By combining the aforementioned conditions, one can isolate the equation that determines the radius of the photon sphere, $r_{\rm ph}$, which is \cite{Uniyal:2022vdu}
\begin{equation}
\label{21}
\frac{d}{dr}h(r)^2|_{r=r_{\rm ph}}=0. 
\end{equation}
 
For the NLED-Bardeen BH, we have
\begin{equation}
\label{22}
4 g^8+16 g^6 r_{\text{ph}}^2+r_{\text{ph}}^6 \left(9 M \sqrt{g^2+r_{\text{ph}}^2}+2 g^2\right)+r_{\text{ph}}^4 \left(17 g^4-26 g^2 M \sqrt{g^2+r_{\text{ph}}^2}\right)-3 r_{\text{ph}}^8=0.
\end{equation}

\begin{figure}[H]
	\begin{minipage}{1.0\textwidth}
    \centering
		\includegraphics[scale=1.1,angle=0]{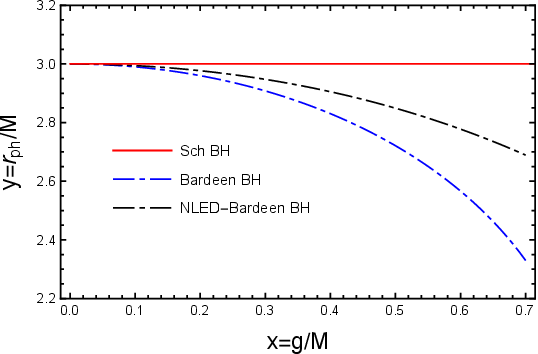}
	\end{minipage}
    \hspace{4cm}
\caption{ The  picture illustrate the dependence of $r_{\rm ph}$ on the magnetic charge $g$.
}\label{Fig.2}
\end{figure}

Due to the highly nonlinear of the resulting algebraic expression for the $r_{\rm ph}$, an exact analytical solution remains intractable. Consequently, we employ numerical root-finding algorithms to determine the functional dependence of $r_{\rm ph}$ on the fundamental spacetime parameters. In Figure \ref{Fig.2}, we illustrate the characteristic evolution of the dimensionless ratio $r_{\rm ph}/M$ as a function of the magnetic charge $g/M$. This numerical mapping allows us to quantify the precise extent to which the NLED-induced corrections shift the critical light orbits relative to the standard Schwarzschild (Sch) and Bardeen cases.

Based on the numerical results presented in Figure \ref{Fig.2}, we perform a comparative analysis of the $r_{\rm ph}$ for three distinct BH configurations. The red solid line represents the standard Sch BH, yielding the well-established constant value of $r_{\rm ph}/M = 3$, which serves as our baseline. The blue line characterizes the Bardeen BH, while the black line accounts for the NLED-Bardeen BH. Our results reveal that the Sch BH possesses the largest $r_{\rm ph}$. For the NLED-Bardeen cases, $r_{\rm ph}/M$ exhibits a monotonic decrease as the magnetic charge $g/M$ increases, suggesting that the presence of magnetic charge enhances the effective gravitational potential and shifts the critical light orbits inward. Most notably, the $r_{\rm ph}$ of the NLED-Bardeen BH (black line) remains consistently larger than that of the Bardeen case (blue line) for a given magnetic charge. Over the displayed parameter range, the numerical solutions satisfy $r_{\rm ph}^{\rm Bardeen}<r_{\rm ph}^{\rm NLED\text{-}Bardeen}<r_{\rm ph}^{\rm Sch}=3M$.


The inward displacement of the photon sphere in the Bardeen spacetime originates from the magnetic charge induced regularization of the gravitational field. Its mass function, $m(r)=Mr^{3}/(r^{2}+g^{2})^{3/2}$, satisfies $m(r)<M$ for finite $r$ and nonzero $g$. Consequently, the gravitational attraction in the photon region is effectively weaker than that in the Schwarzschild geometry with the same asymptotic mass. The maximum of the null effective potential is therefore shifted inward, yielding $r_{\rm ph}^{\rm Bardeen}<3M$. This effect becomes increasingly pronounced as $g/M$ grows.

For the NLED-Bardeen BH, the background metric function $f(r)$ remains identical to that of the Bardeen solution; hence, the magnetic charge induced inward shift persists. However, NLED modifies the effective optical geometry experienced by photons. In particular, the angular function becomes $C(r)=2r^{2}(r^{2}+g^{2})/(3r^{2}-4g^{2})$, and the photon sphere condition is changed. For the Bardeen BH, $h_{\rm Bardeen}^{2}(r)=r^{2}/f(r)$, whereas the NLED-Bardeen case is described by $h_{\rm NLED}^{2}(r)=Q(r)h_{\rm Bardeen}^{2}(r)$, with $Q(r)=2(r^{2}+g^{2})/(3r^{2}-4g^{2})$. In the physically relevant region $r>r_{\star}=2g/\sqrt{3}$, one has $Q'(r)=-28rg^{2}/(3r^{2}-4g^{2})^{2}<0$. Since $d h_{\rm Bardeen}^{2}/dr=0$ at the Bardeen photon sphere, it follows that $d h_{\rm NLED}^{2}/dr<0$ at the same radius. As the outer unstable photon orbit is determined by the minimum of $h^{2}(r)$, the minimum of $h_{\rm NLED}^{2}(r)$ is shifted to a larger radius, yielding $r_{\rm ph}^{\rm NLED-Bardeen}>r_{\rm ph}^{\rm Bardeen}$. Physically, the NLED correction modifies the effective light-cone structure and optical geometry experienced by nonradial photons, thereby partially compensating for the inward shift induced by the Bardeen magnetic charge. Nevertheless, because the background metric remains that of the Bardeen spacetime, the regularizing effect of the magnetic charge still weakens the effective gravitational attraction relative to the Sch case.


To quantify the observable dimensions of the BH, we must establish the definition of the shadow radius $R_{\rm sh}$. This is achieved by considering light rays traced backwards in time from a static observer positioned at a finite or asymptotic radial coordinate $r_\mathrm{o}$. The angular $\alpha$, subtended between the incoming light ray and the local radial direction at the observer's location, is determined by the following geometric relation \cite{Perlick:2021aok,Perlick:2015vta}
\begin{equation}
\label{23}
\cot \alpha=\left.\frac{\sqrt{B(r)}}{\sqrt{C(r)}} \frac{d r}{d \phi}\right|_{r=r_{\mathrm{o}}}.
\end{equation}

Subsequently, by employing Eq. (\ref{19}), we obtain
\begin{equation}
 \left\{\begin{split}
&\cot ^2 \alpha=\frac{h\left(r_{\mathrm{o}}\right)^2}{h(\bar{r})^2}-1, \\
&\sin ^2 \alpha=\frac{h(\bar{r})^2}{h\left(r_{\mathrm{o}}\right)^2},\\
\end{split}\right. \label{24}
\end{equation}
here $\bar{r}$ denotes the turning point of the trajectory, representing the minimum radial distance reached by a light ray before receding to infinity. The boundary of the BH shadow is defined by the set of critical trajectories that, when traced backward in time, asymptotically toward the unstable circular orbit at $r = r_{\text{ph}}$. Consequently, the observed angular radius of the shadow, $\alpha_{\text{sh}}$, is determined by
\begin{equation}
  \sin ^2 \alpha_{\mathrm{sh}}=\frac{h\left(r_{\mathrm{ph}}\right)^2}{h\left(r_{\mathrm{o}}\right)^2}.\label{25} 
\end{equation}

Since the critical impact parameter $b_{\rm c}$ determines the apparent silhouette of the BH for a distant observer and is related to the photon sphere radius by $b_{\rm c}=h(r_{\rm ph})$ \cite{Perlick:2021aok,Perlick:2015vta}, Eq. \eqref{25} can be further simplified as
\begin{equation}
 \label{26} 
\sin ^2 \alpha_{\mathrm{sh}}=\frac{b_{\mathrm{c}}^2}{h\left(r_{\mathrm{o}}\right)^2}, \quad \text { or } \quad \sin ^2 \alpha_{\mathrm{sh}}=\frac{b_{\mathrm{c}}^2 A\left(r_{\mathrm{o}}\right)}{C\left(r_{\mathrm{o}}\right)} .
\end{equation}

For asymptotically flat BHs, it is well established that the shadow radius observed at spatial infinity is equal to the critical impact parameter, namely $R_{\rm sh}=b_{\rm c}$ \cite{Perlick:2021aok}. In contrast, the effective metric in Eq. \eqref{14}, which arises from the NLED correction, does not satisfy the standard asymptotically flat condition. Nevertheless, by evaluating the shadow radius in the local orthonormal frame of an observer located at a sufficiently large radius $r_{\mathrm{o}}$, we find that $R_{\rm sh}$ asymptotically approaches $\sqrt{3/2}b_{\rm c}$. This additional factor originates from the asymptotic behavior of the effective metric function, $A(r_\mathrm{O})/C(r_\mathrm{o})\rightarrow 3/2r^2_{\mathrm{o}}$ as $r_{\mathrm{o}}\rightarrow\infty$. In the following analysis, we adopt $R_{\rm sh}=\sqrt{3/2}b_{\rm c}$ to investigate the relevant optical observables and to distinguish the NLED-Bardeen BH from the Bardeen solution.

\subsection{Verification of the Horizon–Photon Sphere–Shadow Inequality for the NLED-Bardeen Black Hole}

For spherical symmetric BHs, the characterization of the BH size is significantly simplified, since the event horizon, photon sphere, and shadow are all spherical geometric objects and can therefore be described by their corresponding radii. This simplicity allows one to formulate universal relations among these characteristic length scales. In particular, assuming the WEC together with a negative trace of the energy-momentum tensor, Hod proved an upper bound on the physical photon sphere radius of BH with ADM mass \(M\) \cite{Hod:2013jhd},
\begin{equation}
\label{27}
R_{\rm ph}\leq 3M .
\end{equation}

Here \(R_{\rm ph}\) denotes the physical radius of the photon sphere, rather than the coordinate location \(r_{\rm ph}\) of the circular photon orbit. For the NLED-Bardeen BH, the relevant radial coordinate is the areal radius defined with respect to the effective optical geometry, which is given by
\begin{equation}
\label{28}
R_{\rm ph}\equiv \sqrt{C(r_{\rm ph})}.
\end{equation}

In the limit \(C(r)\to r^2\), the physical radius reduces to the coordinate radius, namely \(R_{\rm ph}=r_{\rm ph}\). Under the same assumptions, Cvetič, Gibbons, and Pope further derived an inequality relating the physical photon sphere radius to the shadow radius \cite{Cvetic:2016bxi},
\begin{equation}
\label{29}
R_{\rm ph}\leq \frac{R_{\rm sh}}{\sqrt{3}} .
\end{equation}

While \(R_{\rm ph}\) and \(R_{\rm sh}\) characterize the apparent size of a spherical symmetric BH through its photon sphere and shadow. Additionally, the intrinsic size of the BH is determined by the physical horizon radius \(R_+\equiv \sqrt{C(r_+)}\). In this context, Penrose's entropy bound can be equivalently expressed as the Riemannian Penrose inequality
\begin{equation}
\label{30}
R_+\leq 2M,
\end{equation}
which is known to hold, in the Riemannian setting, under the dominant energy condition. Together with the photon sphere-shadow relation and the horizon bound considered in the literature, this motivates the chain of inequalities \cite{Lu:2019zxb}
\be
\label{31}
\frac{3}{2}R_+\leq R_{\rm ph} \leq \frac{R_{\rm sh}}{\sqrt{3}} \leq 3M.
\ee

This chain of inequalities has been examined for a broad class of BHs satisfying at least the null energy condition, and no counterexample has been identified to date. Hod's original bound was formulated in terms of the innermost photon sphere. For most static and spherical symmetric BHs, however, only a single photon sphere exists, and the distinction between the innermost and outermost photon spheres is therefore irrelevant. In spacetimes admitting multiple photon spheres, the outermost photon sphere is of greater observational relevance because it determines the boundary of the shadow perceived by a distant observer. Accordingly, the inequality chain is understood to refer to the outermost photon sphere and its associated shadow.

Recent studies have further demonstrated that these inequalities can be established under appropriate energy conditions; in particular, Hod's bound requires only the null energy condition. The Bardeen BH satisfies the WEC but violates the dominant energy condition, while the trace of its energy-momentum tensor is positive. Although these properties lie outside the assumptions of Theorems \eqref{27} and \eqref{29}, the complete sequence of inequalities in \eqref{31} has nevertheless been shown to hold for the Bardeen BH \cite{Lu:2019zxb}. In the NLED-Bardeen case, however, photons do not follow the null geodesics of the background spacetime metric, but instead propagate along the null geodesics of the effective metric induced by NLED. Therefore, the relevant quantities for testing the above hierarchy are the physical horizon radius, the physical photon sphere radius, and the shadow radius. The purpose of this subsection is to verify whether the horizon-photon sphere-shadow inequality continues to hold for the NLED-Bardeen BH when these physical geometric radii are used.

To examine whether the above inequality is satisfied, we introduce three dimensionless quantities, $X$, $Y$, and $Z$, defined as 
\be
\label{32}
X=\frac{27 M^2}{R_{\mathrm{sh}}^2}, \quad Y=\frac{R_{\mathrm{sh}}^2}{3 R_{\mathrm{ph}}^2}, \quad Z=\frac{4 R_{\mathrm{ph}}^2}{9 R_{+}^2} .
\ee

The inequality is therefore equivalent to demonstrating that $X\geq1$, $Y\geq1$, and $Z\geq1$. For a fixed BH mass $M$, larger values of $X$, $Y$, and $Z$ correspond to a more compact BH. After a lengthy calculation, the analytical expressions for the dimensionless quantities $X$, $Y$, and $Z$ can be given by
\begin{equation}
 \left\{\begin{split}
&X=\frac{9 \left(4 g^2-3 r^2_{\rm ph}\right)^2 \left(r^2_{\rm ph}-2 g^2\right)}{r^4_{\rm ph} \left(g^2+r^2_{\rm ph}\right) \left(9 r^2_{\rm ph}-26 g^2\right)},\\
&Y=\frac{9r^4_{\rm ph}-26g^2r^2_{\rm ph}}{16g^4-20g^2r^2_{\rm ph}+6r^4_{\rm ph}},\\
&Z=\frac{4  r_{\text{ph}}^2\left(3 r_+^2-4 g^2\right) \left(g^2+r_{\text{ph}}^2\right)}{9 r_+^2 \left(g^2+r_+^2\right) \left(3 r_{\text{ph}}^2-4 g^2\right)}.
\end{split}\right. \label{33}
\end{equation}

\begin{figure}[htbp]
    \centering

    \begin{minipage}{0.38\textwidth}
        \centering
        \includegraphics[width=\linewidth]{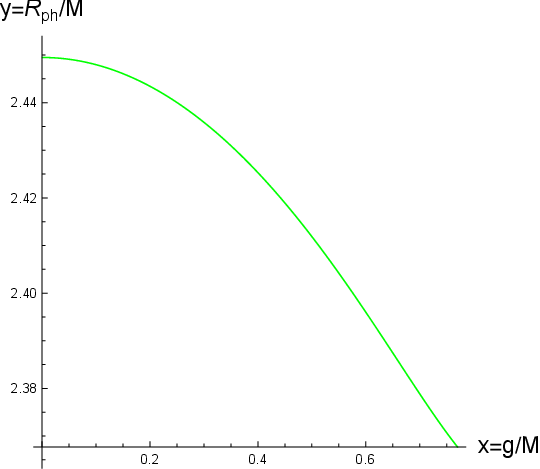}
    \end{minipage}
    \hfill
    \begin{minipage}{0.38\textwidth}
        \centering
        \includegraphics[width=\linewidth]{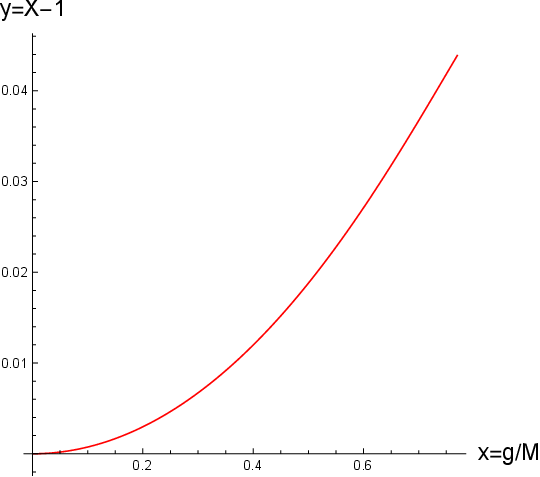}
    \end{minipage}

    \par\medskip

    \begin{minipage}{0.38\textwidth}
        \centering
        \includegraphics[width=\linewidth]{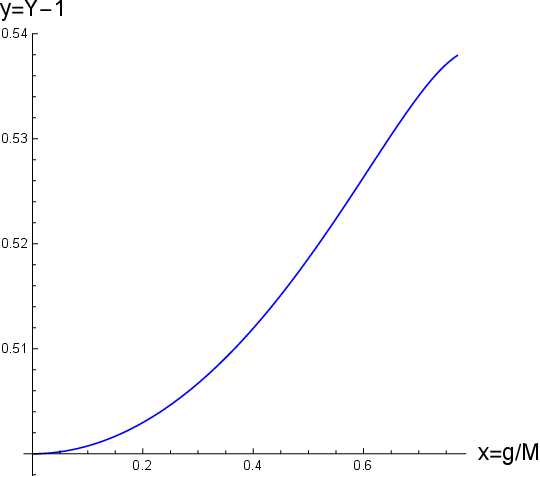}
    \end{minipage}
    \hfill
    \begin{minipage}{0.38\textwidth}
        \centering
        \includegraphics[width=\linewidth]{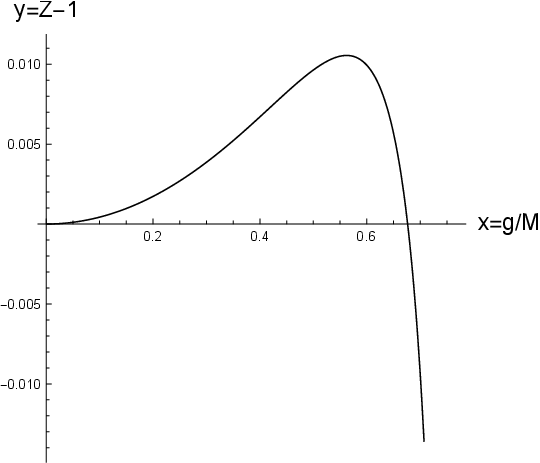}
    \end{minipage}

    \caption{The behavior of the dimensionless functions $X-1$, $Y-1$, $Z-1$, and $R_{\rm ph}$ as functions of the magnetic charge $g/M$.}
    \label{Fig.3}
\end{figure}

Because the analytical expressions are highly involved, we examine the auxiliary functions $X-1$, $Y-1$, and $Z-1$ over the parameter interval $0\leq g/M\leq0.7698$ to test the inequality in Eq. \eqref{31}. As shown in Fig. \ref{Fig.3}, $X-1$ is positive and monotonically increasing throughout this interval. Hence, $X>1$ is always satisfied, which is equivalent to the bound $R_{\rm sh}\leq 3\sqrt{3}M$. Additionally, $Y-1$ remains positive and monotonically increasing throughout the physical parameter range, implying that $Y>1$. Therefore, the corresponding inequality in Eq. \eqref{29} is preserved; equivalently, Eq. \eqref{29} is satisfied for all allowed values of the magnetic charge. In contrast, the function $Z-1$ first increases and then decreases as $g/M$ grows. Although the condition $Z\geq1$ is satisfied over most of the parameter range, it does not hold universally. Furthermore, the physical photon sphere radius $R_{\rm ph}$ exhibits behavior consistent with that of the coordinate photon sphere radius $r_{\rm ph}$ and satisfies Hod's bound in Eq.~\eqref{27}, in agreement with the Bardeen case. Nevertheless, since the remaining inequalities are not satisfied over the entire parameter space, the complete horizon--photon sphere-shadow inequality chain is not universally valid for the NLED-Bardeen BH.

The violation of the inequality can be understood from the fact that the assumptions underlying the original horizon-photon sphere-shadow relation are not fully satisfied in the NLED-Bardeen spacetime. This relation is generally derived under the assumption that the event horizon, photon sphere, and shadow are all described by a single asymptotically flat background metric. In the NLED-Bardeen case, however, the event horizon remains determined by the background Bardeen metric, whereas the photon sphere and shadow are governed by the NLED-induced effective optical geometry. Moreover, the effective metric does not exhibit the standard asymptotically flat normalization. Consequently, the shadow radius is no longer directly given by the critical impact parameter $b_{\rm c}$, but instead satisfies $R_{\rm sh}=\sqrt{3/2}b_{\rm c}$ for a sufficiently distant observer. Therefore, NLED modifies the photon light-cone structure, the condition for circular photon orbits, and the observed shadow scale without inducing a corresponding modification of the background horizon. Since these three physical scales are no longer constrained by a common geometric structure, the conventional horizon-photon sphere-shadow inequality chain may be violated.

\subsection{Categories of photon trajectories: Direct emission, Lensing rings, and Photon rings}
To delineate the observational signatures of BH \eqref{14}, we characterize its photon and lensing ring morphologies within the framework of an optically thick accretion disk \cite{Uniyal:2022vdu,Peng:2020wun,Gralla:2019xty}. This investigation is predicated on a rigorous treatment of the null geodesics that govern light propagation. By laying the theoretical groundwork through a systematic analysis of photon trajectories in the strong-field regime, the general geodesic relations established in Eq. \eqref{19} are cast into the following explicit form
\be
\label{34}
\left(\frac{d u}{d \phi}\right)^2=u^4 C^2\left(\frac{1}{u^2}\right)\left[  \frac{1}{b^2 A\left(\frac{1}{u}\right)B\left(\frac{1}{u}\right)}-\frac{1}{C\left(\frac{1}{u}\right)B\left(\frac{1}{u}\right)}\right]\equiv G(u),
\ee  
where we define \( u \equiv 1/r \) and \( b \equiv \omega_0 / p_{\phi} \) in order to express the photon trajectories in a more transparent and tractable form. Focusing on the equatorial slice ($\theta = \pi/2$), the null geodesics are fundamentally categorized by the critical impact parameter $b_c$, which dictates the ultimate fate of the incident photons \cite{Gralla:2019xty}. Depending on the value of $b$ relative to this threshold, the trajectories manifest in three distinct dynamical regimes: Scattering Regime ($b > b_c$): Photons originating from the asymptotic far-field are gravitationally deflected. These trajectories reach a periapsis (minimum radial distance) $r_{\text{min}}$ before escaping back to infinity, forming the basis of gravitational lensing. Capture Regime ($b < b_c$): The gravitational attraction overwhelms the centrifugal barrier, causing photons to cross the event horizon. This inward flux defines the photon capture cross-section of the BH. Critical Regime ($b = b_c$): These trajectories represent the unstable equilibrium state where photons asymptotically spiral toward the photon sphere at $r_{\rm ph}$. This critical set of geodesics delineates the apparent boundary, or the BH shadow.

The total azimuthal deflection $\phi$ accumulated along a photon's trajectory is critically determined by the impact parameter $b$. For the capture regime ($b < b_c$), the light ray terminates at the event horizon $u_{\rm H} = 1/r_{\rm +}$, while for the scattering regime ($b > b_c$), the trajectory is symmetric about the periapsis $u_{\min}$, defined by the largest positive root of the orbital equation $G(u) = 0$. These integrated paths are expressed as

\begin{equation}
 \left\{\begin{split}
&\phi=\int_0^{u_{\rm H}} \frac{d u}{\sqrt{G(u)}}, \quad b<b_{c}, \\
&\phi=2 \int_0^{u_{\rm m}} \frac{d u}{\sqrt{G(u)}}, \quad b>b_{c}.\\
\end{split}\right. \label{35}
\end{equation}

To disentangle the complex radiative signatures emerging from the strong-field regime, we adopt the tripartite classification of photon trajectories—comprising direct emission, the lensing ring, and the photon ring—as first formalized in Refs. \cite{Gralla:2019xty}. This taxonomic framework maps the cumulative deflection of light rays to specific image sub-structures, where each component is distinguished by the number of times the trajectory intersects the accretion disk before reaching the observer. Based on the fractional orbit number $n(b) = \phi / 2\pi$. This quantity represents the number of full revolutions performed by the photon before reaching the observer. As shown in Refs. \cite{Yang:2022btw,Peng:2020wun}, this quantity is discretely quantized, and its explicit dependence on the $b$ can be written as 
\begin{equation}
 \label{36}
n(b) = \frac{2m - 1}{4}, \quad m \in \mathbb{Z}^+.
\end{equation}

The transition between different trajectory families is marked by critical thresholds $b_m^{\pm}$, which partition the impact parameter space into three distinct observational components based on their interaction with the accretion disk:
\begin{itemize}
\item Direct Emission ($1/4 < n < 3/4$): These trajectories, spanning $b \in (0, b_2^-) \cup (b_2^+, \infty)$, intersect the disk plane only once. 

\item Lensing Ring ($3/4 < n < 5/4$): Comprising rays in the range $b\in (b_2^-, b_3^-) \cup (b_3^+, b_2^+)$, these photons undergo significant gravitational bending, resulting in exactly two intersections with the disk.

\item Photon Ring ($n > 5/4$): Restricted to the narrow interval $b \in (b_3^-, b_3^+)$, these photons execute multiple revolutions ($n \ge 3$ disk crossings). 
\end{itemize}

\begin{figure}[h]
\centering
	\begin{minipage}{0.3\textwidth}
    \centering
		\includegraphics[scale=0.5,angle=0]{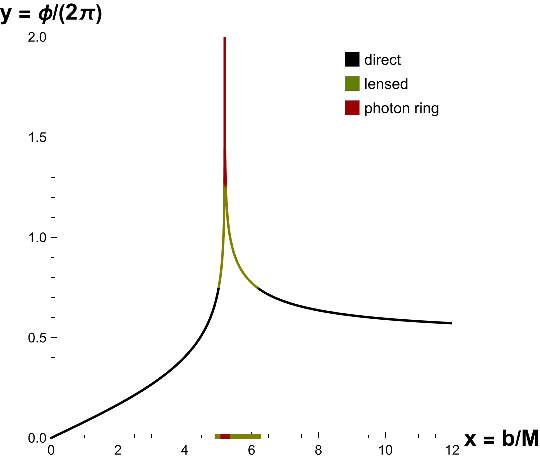}
	\end{minipage}
    \begin{minipage}{0.3\textwidth}
    \centering
 		\includegraphics[scale=0.5,angle=0]{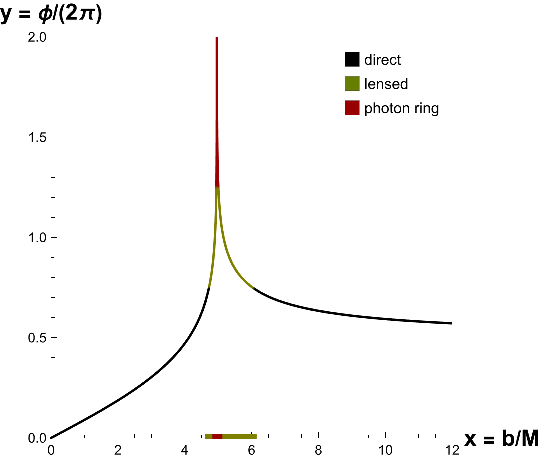}
	\end{minipage}	
    \begin{minipage}{0.3\textwidth}
    \centering
		\includegraphics[scale=0.5,angle=0]{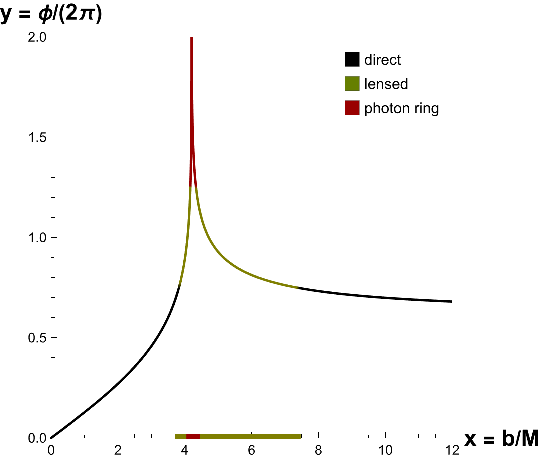}
	\end{minipage}
  \begin{minipage}{0.3\textwidth}
    \centering
		\includegraphics[scale=0.5,angle=0]{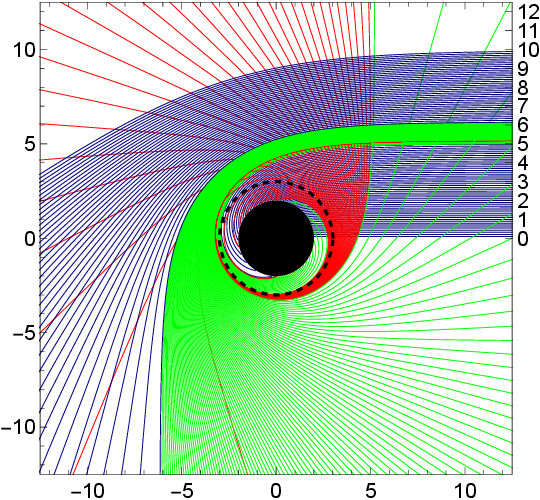}
        \smallskip
        \textit{\rm (a) Sch BH }
	\end{minipage}
    \begin{minipage}{0.3\textwidth}
    \centering
		\includegraphics[scale=0.5,angle=0]{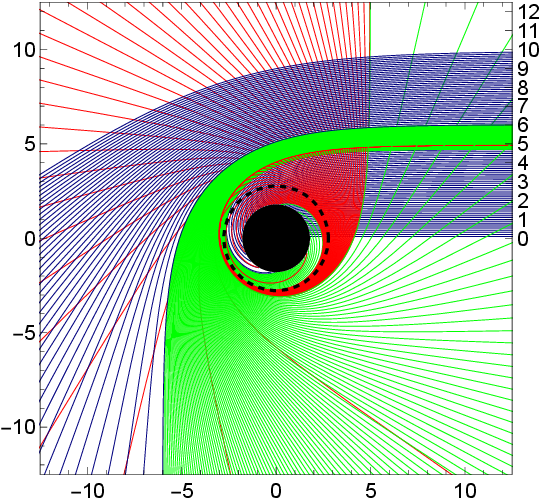}
        \smallskip
        \textit{\rm (b) Bardeen BH }
	\end{minipage}
    \begin{minipage}{0.3\textwidth}
    \centering
		\includegraphics[scale=0.5,angle=0]{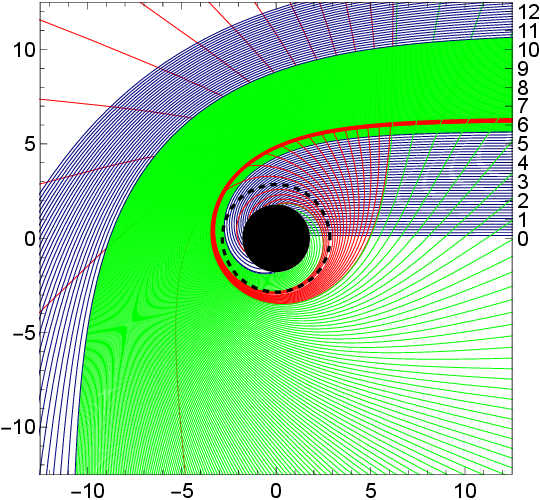}
        \smallskip
      \textit{\rm (c) NLED-Bardeen BH }
    \end{minipage}	
\caption{ Figure illustrates the parametric evolution of the winding number $n$ alongside the corresponding equatorial trajectories for Sch ($g/M=0$), Bardeen ($g/M=0.5$), and NLED-Bardeen ($g/M=0.5$) BHs. The upper panels delineate the transition between distinct image sub-structures-specifically direct emission, lensing rings, and photon rings—as categorized by the thresholds of the winding number $n = \phi/2\pi$ ($n < 3/4$ in black, $3/4 < n < 5/4$ in green, and $n > 5/4$ in red). Complementing this classification, the lower panels map these geodesics in the $(r, \phi)$ plane, where the central event horizon and the critical photon sphere are represented by a filled disk and a dashed circle, respectively. } \label{Fig.4}
\end{figure}

The photon dynamics in different BHs are illustrated in Fig. \ref{Fig.4}. The upper panels display the total number of orbital windings, while the lower panels show the corresponding photon trajectories around the BHs. In the first row, the dependence of the orbital winding number on the impact parameter is presented for photons propagating near a thin accretion disk: the red curves correspond to photon ring trajectories, the green curves to lensing ring trajectories, and the black curves to direct emission. The second row depicts the associated light paths, where direct emission is shown in blue, lensing ring trajectories in green, and photon ring trajectories in red, providing a complete classification of photon motion in these spacetimes.

To provide a quantitative comparison of the influence of BH parameters, the characteristic values for the event horizon, photon sphere, and shadow radius are summarized in Tables \ref{Tab.1} and \ref{Tab.2}. From Table \ref{Tab.1}, the Sch BH gives the reference values $r_{\rm +}=2M$ and $r_{\rm ph}=3M$. For the Bardeen-type solutions with $g/M=0.5$, the outer horizon is smaller than the Schwarzschild value because the magnetic charge regularizes the central geometry through the mass function $m(r)=Mr^3/(r^2+g^2)^{3/2}$. Since $m(r)<M$ at finite radius, the effective gravitational attraction in the photon region is weaker than that of the Schwarzschild spacetime with the same ADM mass, and the characteristic radii are shifted inward.

A crucial distinction, however, appears between the  Bardeen BH and the NLED-Bardeen BH. For the same magnetic charge, both solutions possess the same event horizon $r_{\rm +}$. This is because the NLED effective metric describes the propagation of electromagnetic perturbations rather than a backreaction-induced modification of the background geometry. In contrast, the photon sphere and shadow are optical quantities; they are governed not by the background metric alone, but by the NLED-induced effective geometry. More precisely, the circular photon orbit is determined by the extremum condition of $h^2(r)=C(r)/A(r)$. In the Bardeen case one has $h^2(r)=r^2/f(r)$, whereas in the NLED-Bardeen case the function is modified by an additional factor originating from the nonlinear electromagnetic Lagrangian. This changes the light-cone structure experienced by photons and shifts the unstable circular orbit relative to the Bardeen geometry.

This explains the hierarchy displayed in Table \ref{Tab.1}. 
For the two asymptotically flat cases, namely the Sch and Bardeen BHs, the shadow radius measured by a distant observer is simply $R_{\rm sh}=b_c$. For the NLED-Bardeen BH, however, the effective metric does not approach the standard asymptotically flat normalization in the optical sector. As a result, the observed shadow scale is not directly given by $b_c$, but instead satisfies $R_{\rm sh}/M=\sqrt{3/2}\, b_c \approx 5.14799$. Therefore, although the critical impact parameter itself is reduced by the NLED correction, the additional factor enlarges the observed shadow radius. This shows that NLED affects not only the location of the photon sphere, but also the mapping between the critical photon orbit and the image seen by a distant observer.

Table \ref{Tab.2} further explores the sensitivity of these physical parameters to the magnetic charge $g$ in the NLED-Bardeen spacetime. As $g/M$ increases, the event horizon $r_{\rm +}$ decreases monotonically because the regularized mass function becomes more strongly suppressed near the central region. The photon sphere radius $r_{\rm ph}$ also decreases, indicating that the unstable circular photon orbit moves closer to the compact object as the magnetic charge grows. Physically, the magnetic charge introduces a regularization scale that softens the central geometry and redistributes the effective gravitational mass; consequently, the characteristic surfaces associated with the horizon and photon trapping are pulled inward.

The behavior of the shadow radius contains an additional optical effect. Since the NLED correction modifies the effective angular component of the metric, photons propagate as if they were moving in a nonlinear optical medium generated by the background electromagnetic field. Increasing $g$ strengthens this optical-medium effect and changes the bending of nonradial photon trajectories. The monotonic decrease of $R_{\rm sh}$ with $g$ therefore reflects the combined influence of two mechanisms: the inward shift of the photon sphere due to the Bardeen magnetic charge and the NLED-induced modification of the effective light cone. In this sense, the NLED-Bardeen BH differs from an ordinary geometrical deformation of spacetime: the horizon remains controlled by the background metric, whereas the photon sphere, critical impact parameter, lensing rings, photon rings, and observed shadow are controlled by the effective optical geometry. This separation is the fundamental reason why NLED can change the observable photon dynamics without producing a corresponding change in the event horizon.

\begin{table*}[h]
\centering
\caption{\label{Tab.1} Table summarizes the dimensionless physical quantities for the three BHs. }
\setlength{\tabcolsep}{2pt} 
\renewcommand{\arraystretch}{1.05} 
\begin{tabular}{|l|cccccccc|}
\hline
BHs & $r_{\rm +}/M$ & $r_{\rm ph}/M$& $b_{c}/M$ & $b^{-}_{1}/M$ &
$b^{-}_{2}/M$  & $b^{+}_{2}/M$ & $b^{-}_{3}/M$& $b^{+}_{3}/M$  \\
\hline
Sch BH $(g/M=0)$
& 2.00000 & 3.00000 & 5.19615 & 2.84770 & 5.01514 & 6.16757 & 5.18781 & 5.22794 \\
\hline
Bardeen BH $(g/M=0.5)$
&1.78597&2.76871 &4.96004 & 2.57082&4.72406 &6.03088 &4.94576&5.00416\\
\hline
NLED-Bardeen BH $(g/M=0.5)$
&1.78597&2.84911 &4.20332&1.86759&3.82378 &7.34842 & 4.16514 &4.33512\\
\hline
\end{tabular}
\end{table*}

\begin{table*}[h]
\centering
\caption{\label{Tab.2}The table summarizes the values of the relevant physical quantities for different choices of the magnetic charge \( g \) of the NLED-Bardeen BH.
  }
\setlength{\tabcolsep}{2pt} 
\renewcommand{\arraystretch}{1.05} 
\begin{tabular}{|l|cccccccc|}
\hline
NLED-Bardeen BH & $r_{\rm +}/M$ & $r_{\rm ph}/M$& $b_{c}/M$ & $b^{-}_{1}/M$ &
$b^{-}_{2}/M$  & $b^{+}_{2}/M$ & $b^{-}_{3}/M$& $b^{+}_{3}/M$  \\
\hline
$g/M=0.1$
& 1.99247& 2.99425 & 4.24107 & 1.95645 & 3.91174 & 6.82170 &4.21232 & 4.35618 \\
\hline
$g/M=0.3$
&1.92962&2.94742 &4.22848 & 1.92951&3.88484 &7.31375&4.19714&4.34875\\
\hline
$g/M=0.5$
&1.78597&2.84911 &4.20332&1.86759&3.82378 &7.34842 & 4.16514 &4.33512\\
\hline
$g/M=0.7$
&1.47436&2.68913 &4.16683&1.73232&3.69432 &7.39610& 4.10934 &4.31906\\
\hline
\end{tabular}
\end{table*}

\section{Observed optical appearance of NLED–Bardeen black hole surrounded by thin disk accretion }
In this section, we characterize the optical appearance of the NLED-Bardeen BH illuminated by a geometrically thin accretion disk \cite{Uniyal:2022vdu,Cai:2025pan,Wu:2025hcu}. We consider a face-on viewing geometry, where a distant observer is positioned along the polar axis. To quantify the observed radiation, we utilize relativistic transfer functions that map local emission from the equatorial accretion flow to the observer’s image plane. These functions encapsulate the full range of GR effects—including gravitational redshift, null geodesic bending, and multiple-orbit lensing—effectively bridging the gap between local emissivity and the observed specific intensity. By integrating photon trajectories weighted by these transfer relations, we can systematically reconstruct the brightness distribution and identify the discrete signatures of the NLED-Bardeen BH.

The specific intensity perceived by a distant observer is governed by the relativistic transfer relation
\begin{equation}
\label{37}
I_o(r, \nu_o) = A(r)^{3/2} I_e(r, \nu_e),
\end{equation}
where $I_e(r, \nu_e)$ and $\nu_e$ represent the local specific intensity and emission frequency at the disk surface, while $I_o(r, \nu_o)$ and $\nu_o$ are the corresponding quantities measured by the polar observer. This mapping is rooted in Liouville’s theorem, which guarantees that the phasespace density, $I_e(r, \nu_e) / \nu^3_e$, remains invariant along the null geodesics. To determine the total observed intensity $I_{\text{obs}}(r)$ at a given radius, we integrate the specific intensity over the entire frequency spectrum
\be
\label{38}
I_{\rm obs}(r)=\int I_o\left(r, v_o\right) d v_o=A(r)^2 I_{\rm e m}(r),
\ee
where the $I_{\text{em}}(r)$ is defined by integrating the local specific intensity over the entire frequency spectrum
\begin{equation}
\label{39}
I_{\text{em}}(r) = \int I_e(r, \nu_e) d\nu_e.
\end{equation}

Assuming a transparent medium where absorption and scattering are neglected, the total observed intensity $I_{\text{obs}}$ is determined by the cumulative emission gathered at each discrete intersection of the photon trajectory with the accretion disk. The number of such crossings is dictated by the geodesic curvature: direct emission intersect the disk once, while lensed and photon ring trajectories undergo two or more crossings as they wrap around the compact object. Consequently, the brightness at any given impact parameter is the superposition of these individual contributions, expressed as
\be
\label{40}
I_{\rm o b s}(b)=\left.\sum_m A(r)^2 I_{\rm e m}(r)\right|_{r=r_m(b)},
\ee
here the transfer function \( r_m(b) \) relates the photon impact parameter to the radial position where the photon intersects the thin disk. Its derivative \( \mathrm{d}r / \mathrm{d}b \), known as the demagnification factor, characterizes the compression or stretching of this mapping \cite{Gralla:2019xty}. A steeper transfer function corresponds to stronger demagnification, which reduces its contribution to the total observed intensity of the BH. Although photons near the photon ring may cross the disk multiple times, the extreme steepness of their transfer functions makes their intensity contribution negligible. Therefore, we consider only the first three intersections \( m = 1, 2, 3 \), with transfer functions given by
\begin{equation}
\left\{\begin{split}
\label{41}
&r_1(b)=\frac{1}{u\left(\frac{\pi}{2}, b\right)}, & b \in\left(b_1^{-}, \infty\right),\\
&r_2(b)=\frac{1}{u\left(\frac{3 \pi}{2}, b\right)}, & b \in\left(b_2^{-}, b_2^{+}\right),\\
&r_3(b)=\frac{1}{u\left(\frac{5 \pi}{2}, b\right)}, & b \in\left(b_3^{-}, b_3^{+}\right).
\end{split}\right.  
\end{equation}

Building upon the taxonomy established in Ref. \cite{Gralla:2019xty}, the observed intensity can be decomposed into a tripartite hierarchy of morphological signatures:
\begin{itemize}
\item Direct emission feature: The first transfer function characterizes the direct image, representing the standard projection of the accretion disk.

\item Lensing ring feature: The second function maps emission from the posterior (far) side of the disk into a secondary, significantly demagnified ring.

\item Photon Ring feature: The third contribution manifests as an attenuated photon ring sourced from the anterior (near) portion of the disk, appearing as a highly compressed innermost feature.
\end{itemize}

\begin{figure}[H]
\centering
	\begin{minipage}{0.3\textwidth}
    \centering
		\includegraphics[scale=0.55,angle=0]{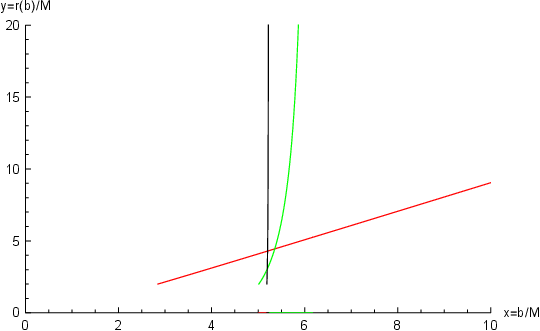}
        \smallskip
        \textit{\rm (a) Sch BH \\$\tan_{r}\theta=0.980311$}
        \end{minipage}
    \begin{minipage}{0.3\textwidth}
    \centering
 		\includegraphics[scale=0.55,angle=0]{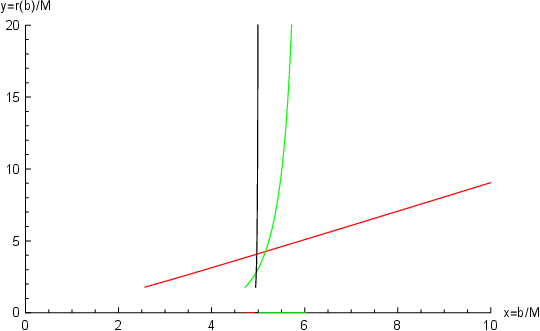}
            \smallskip
        \textit{\rm (b) Bardeen BH 
        \\$\tan_{r}\theta=0.971313$}
	\end{minipage}
    \begin{minipage}{0.3\textwidth}
    \centering
		\includegraphics[scale=0.55,angle=0]{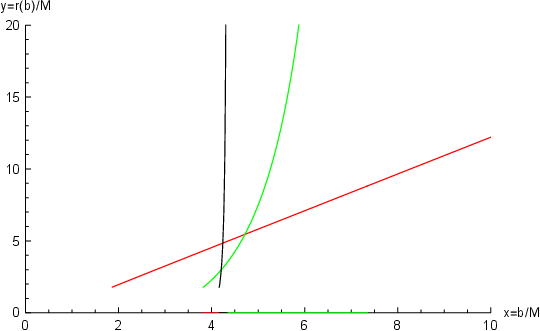}
            \smallskip
        \textit{\rm (c) NLED-Bardeen BH \\$\tan_{r}\theta=1.281370$}
	\end{minipage}

\centering
	\begin{minipage}{0.3\textwidth}
    \centering
		\includegraphics[scale=0.55,angle=0]{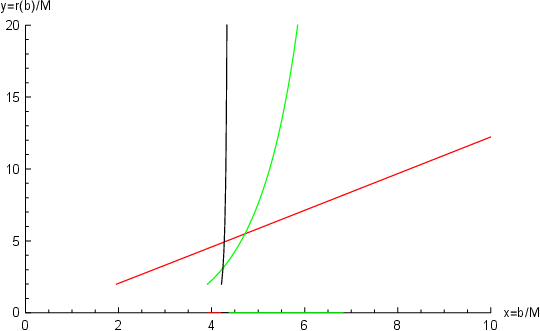}
            \smallskip
        \textit{\rm (d) NLED-Bardeen BH
        \\$\tan_{r}\theta=1.272580$}
	\end{minipage}
    \begin{minipage}{0.3\textwidth}
    \centering
 		\includegraphics[scale=0.55,angle=0]{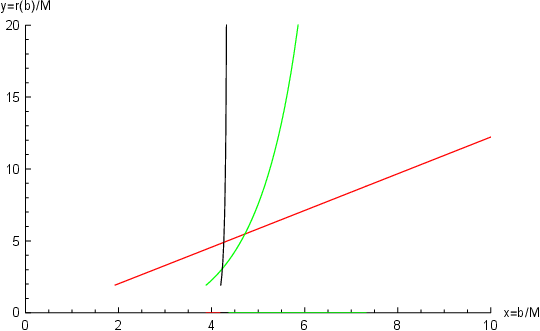}
            \smallskip
        \textit{\rm (e) NLED-Bardeen BH\\
        $\tan_{r}\theta=1.275350$}
	\end{minipage}
    \begin{minipage}{0.3\textwidth}
    \centering
		\includegraphics[scale=0.55,angle=0]{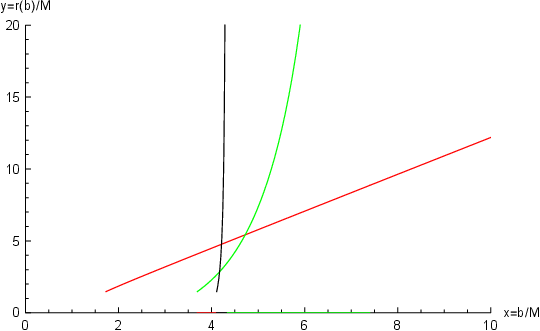}
            \smallskip
        \textit{\rm (f) NLED-Bardeen BH\\
        $\tan_{r}\theta=1.292030$}
	\end{minipage}
\caption{The upper panels present the three transfer functions for the Sch BH, the Bardeen BH with $g/M=0.5$, and the NLED-Bardeen BH with $g/M=0.5$. The lower panels show the transfer functions of the NLED-Bardeen BH for three representative magnetic charges, $g/M=0.1$, $0.3$, and $0.7$. The red, green, and black curves correspond to the first-, second-, and third-order transfer functions, respectively, mapping the direct emission, lensing ring, and photon ring onto the observer's image plane.
 } \label{fig.5}
\end{figure}

The upper three rows of Fig. \ref{fig.5} display the three transfer functions for the Sch BH, the Bardeen BH with $g/M=0.5$, and the NLED-Bardeen BH with $g/M=0.5$. The red, green, and black curves represent the first-, second-, and third-order transfer functions, corresponding to the direct emission, lensing ring, and photon ring contributions. More specifically, the transfer function $r_m(b)$ maps the impact parameter $b$ on the observer's screen to the radial position on the thin accretion disk at which a photon intersects the disk for the $m$-th time. 
For the first-order transfer functions, the corresponding slopes are approximately $\tan_r\theta=0.980311$, $0.971313$, and $1.281370$ for the Sch, Bardeen, and NLED-Bardeen BHs, respectively. This indicates that the Bardeen case provides the largest contribution to the observed total intensity, followed by the Sch case, whereas the NLED-Bardeen case gives the weakest contribution. 
Moreover, the higher-order transfer functions are generally much steeper than the first-order one. Although the second- and third-order images are generated by photons that undergo strong gravitational deflection near the photon sphere, their corresponding disk regions are strongly compressed on the observer's screen. Their flux contributions are therefore highly suppressed, whereas the first-order transfer function, associated with direct emission, provides the dominant contribution to the total observed intensity.

The bottom row of Fig. \ref{fig.5} provides a visual representation of the transfer functions for the NLED-Bardeen BH at three representative values of the magnetic charges, $g/M=0.1$, $0.3$, and $0.7$.
For the three cases considered, the average slopes of the first transfer functions are $1.27258$, $1.27535$, and $1.29203$. The monotonic increase of this slope with the magnetic charge $g$ indicates that the contribution of the direct image to the total observed intensity is gradually suppressed as $g$ increases. Physically, increasing $g$ modifies the NLED-corrected optical geometry through the effective metric felt by photons. As a result, the photon sphere, critical impact parameter, and bending behavior of null trajectories are shifted. The magnetic charge therefore changes not only the background gravitational potential, but also the effective propagation cone of light.

To characterize the optical appearance of the NLED-Bardeen BH, we reconstruct its observational signatures by ray tracing photon trajectories from the strong-field region to the screen of a distant observer. The shadow boundary is determined numerically from critical null geodesics that asymptotically approach the unstable photon sphere and are mapped onto celestial coordinates. This procedure provides a direct visualization of the effects of the magnetic charge on both the shadow morphology and the spatial distribution of the observed brightness.

The observed intensity is then obtained by combining a prescribed radial emissivity profile for the thin accretion disk with the corresponding relativistic transfer functions. These transfer functions encode the mapping between emission points on the disk and positions on the observer's screen, while incorporating the effects of gravitational redshift and strong lensing. To assess the dependence of the optical appearance on the assumed radiation profile, we consider three idealized, time-independent emissivity models. Although these prescriptions do not aim to reproduce the detailed microphysics of realistic accretion disks, they provide a controlled framework for isolating the effects of the spacetime geometry and NLED-modified photon propagation on the resulting image.

In Model 1, the disk is assumed to be truncated at the innermost stable circular orbit $(r_{\rm isco})$. The emission vanishes at and inside this radius, reaches its maximum value immediately outside the $r_{\rm isco}$, and subsequently decreases according to an inverse-square profile:
\be 
\label{42} 
I_{\rm em}\left(r\right)= \begin{cases} I_0\left(\frac{1}{r-\left(r_{\mathrm{isco}}-1\right)}\right)^2, & r>r_{\mathrm{isco}} \\ 0, 
& r \leq r_{\mathrm{isco}}
\end{cases} 
\ee
where $I_0$ denotes the maximum value of the emitted intensity. Furthermore, this model describes a disk whose radiation is concentrated near its inner edge and gradually weakens at larger radii. In Model 2, the emission profile is instead anchored at the photon sphere radius, $r_{\rm ph}$. The intensity is maximal immediately outside $r_{\rm ph}$ and decreases more rapidly, following an inverse-cubic law:
\be 
\label{43} 
I_{\rm e m}(r)= \begin{cases}I_0\left(\frac{1}{r-\left(r_{\rm p h}-1\right)}\right)^3, & r>r_{\rm p h} \\ 0.
& r \leq r_{\rm p h}
\end{cases} 
\ee

Compared with Model 1, this prescription places greater emphasis on radiation originating from the strongly lensed region near the unstable circular photon orbit, thereby providing a simple phenomenological model for a highly centrally concentrated emission profile. Finally, Model 3 allows the disk to emit down to the event horizon, $r_{\rm +}$, without imposing a sharp cutoff at either $r_{\rm ph}$ or $r_{\rm isco}$. The emissivity is normalized to $I_0$ at the horizon and decreases smoothly with increasing radius according to
\be 
\label{44} 
I_{\rm e m}(r)= \begin{cases}I_0 \frac{\frac{\pi}{2}-\arctan \left(r-\left(r_{\rm i s c o}-1\right)\right)}{\frac{\pi}{2}-\arctan \left(r_{\rm +}-\left(r_{\rm i s c o}-1\right)\right)}, & r>r_{\rm +} \\ 0. 
& r \leq r_{\rm +}
\end{cases} 
\ee

This profile produces a continuous and monotonic radial decline in brightness, without a localized emission peak. Taken together, these three prescriptions enable us to assess the robustness of the image features and to distinguish geometric effects from those associated with the assumed disk emissivity.

\begin{figure}[htbp]
\centering
	\begin{minipage}[c]{0.3\textwidth}
	    \centering
		\includegraphics[scale=0.26,angle=0]{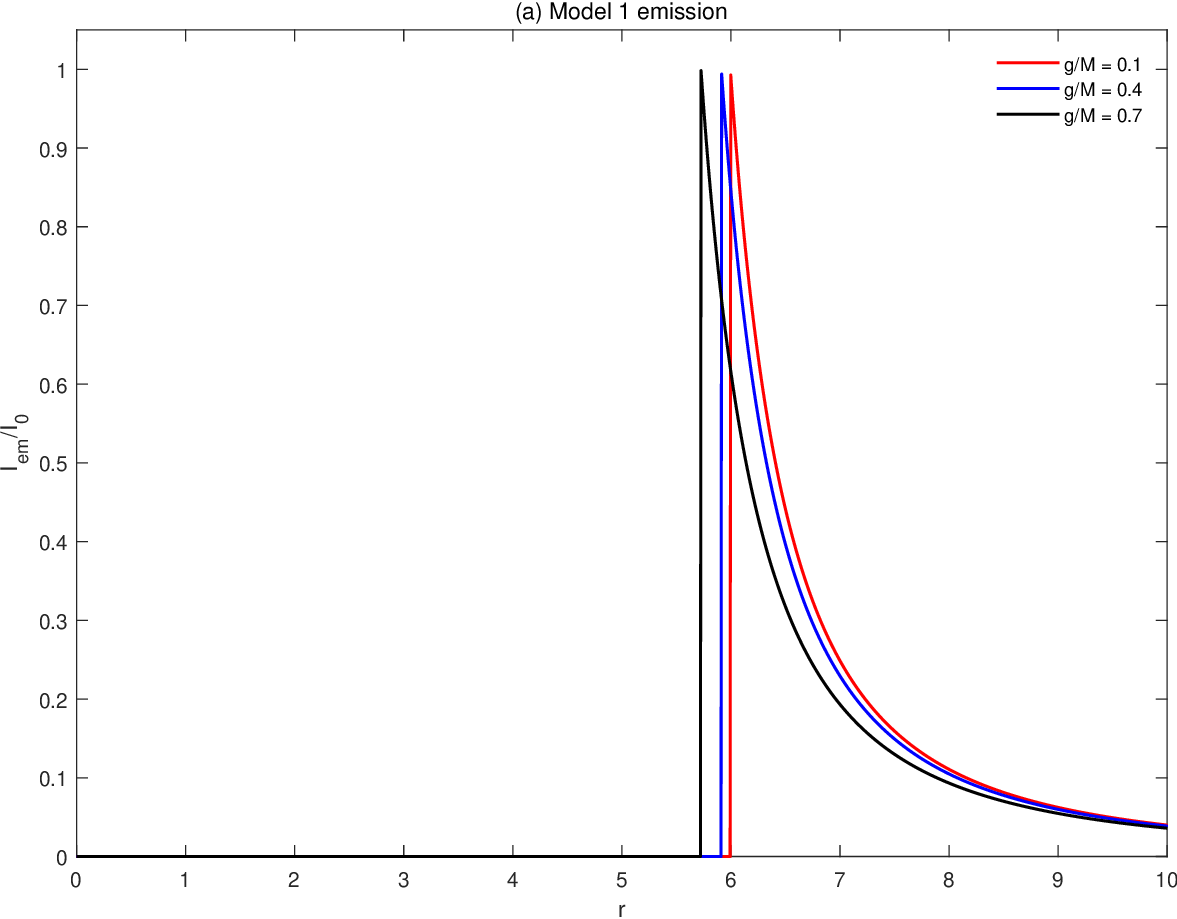}
	\end{minipage}
    \hspace{0.5cm}
    \begin{minipage}[c]{0.3\textwidth}
        \centering
		\includegraphics[scale=0.26,angle=0]{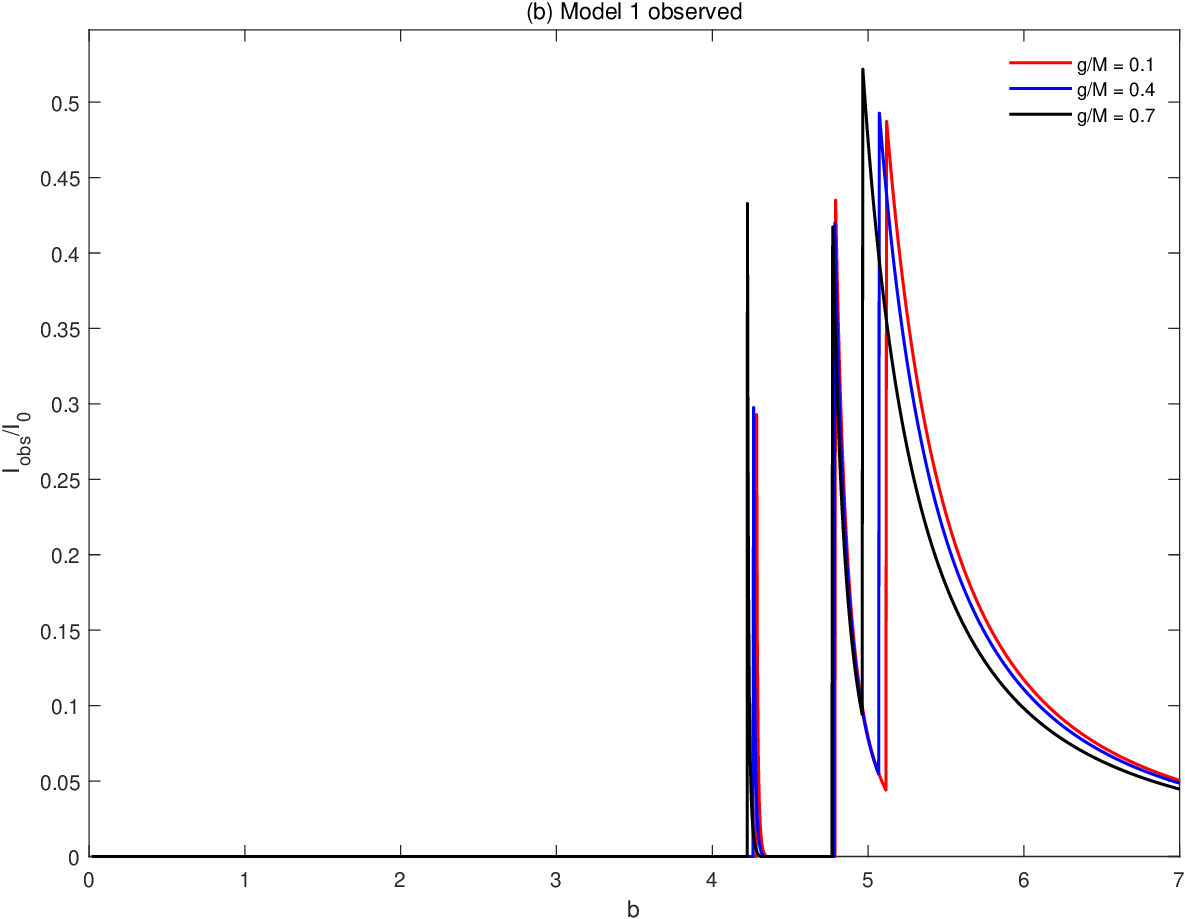}
	\end{minipage}	
   \begin{minipage}[c]{0.34\textwidth}
    \centering
    \includegraphics[height=3.9cm,scale=0.8]{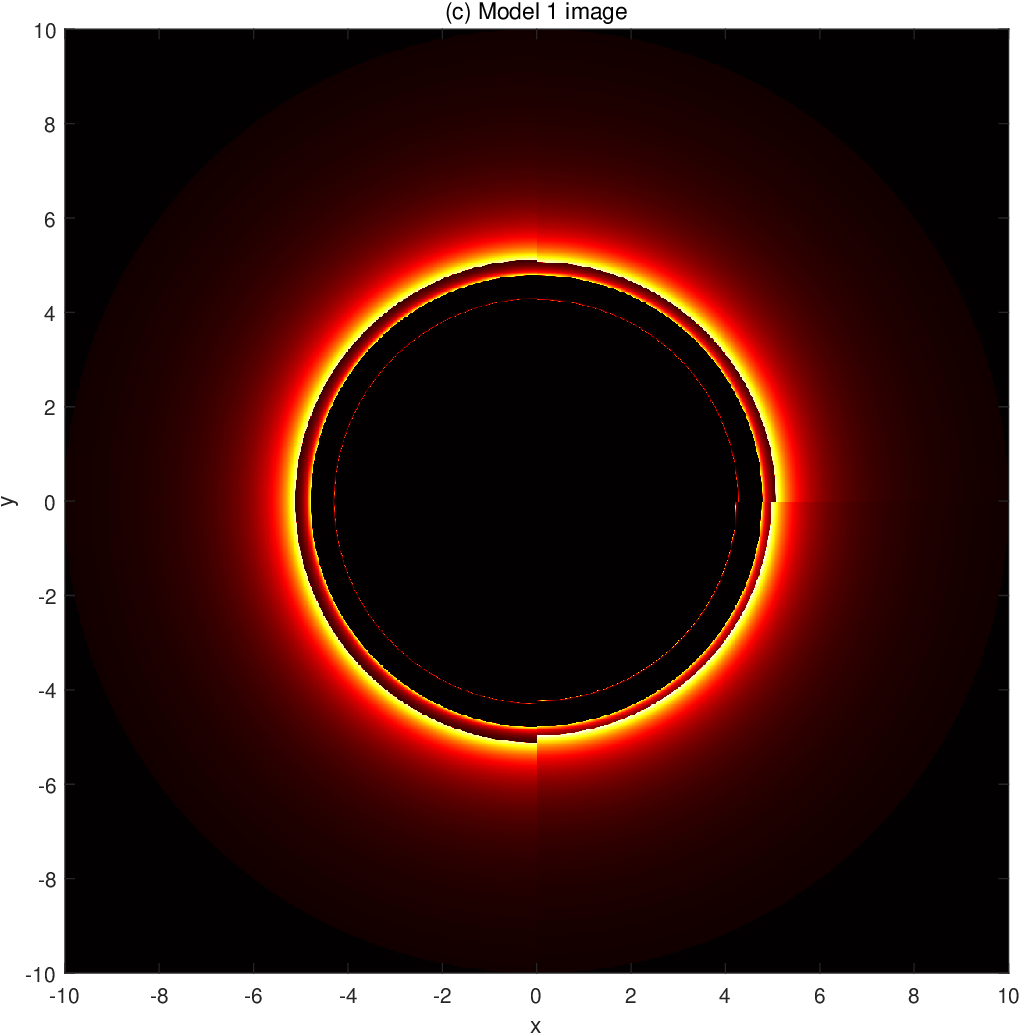}
    \hspace{-0.25cm}
    \raisebox{0cm}{%
        \includegraphics[height=3.9cm,width=0.11\linewidth]{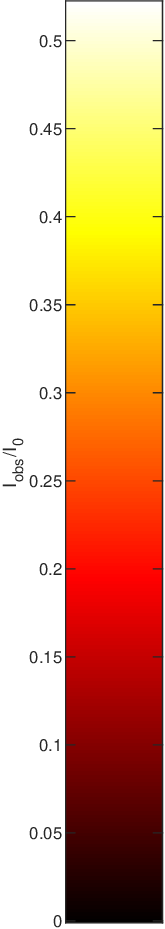}
    }
\end{minipage}

\vspace{2em}

 \centering
	\begin{minipage}[c]{0.3\textwidth}
	    \centering
		\includegraphics[scale=0.26,angle=0]{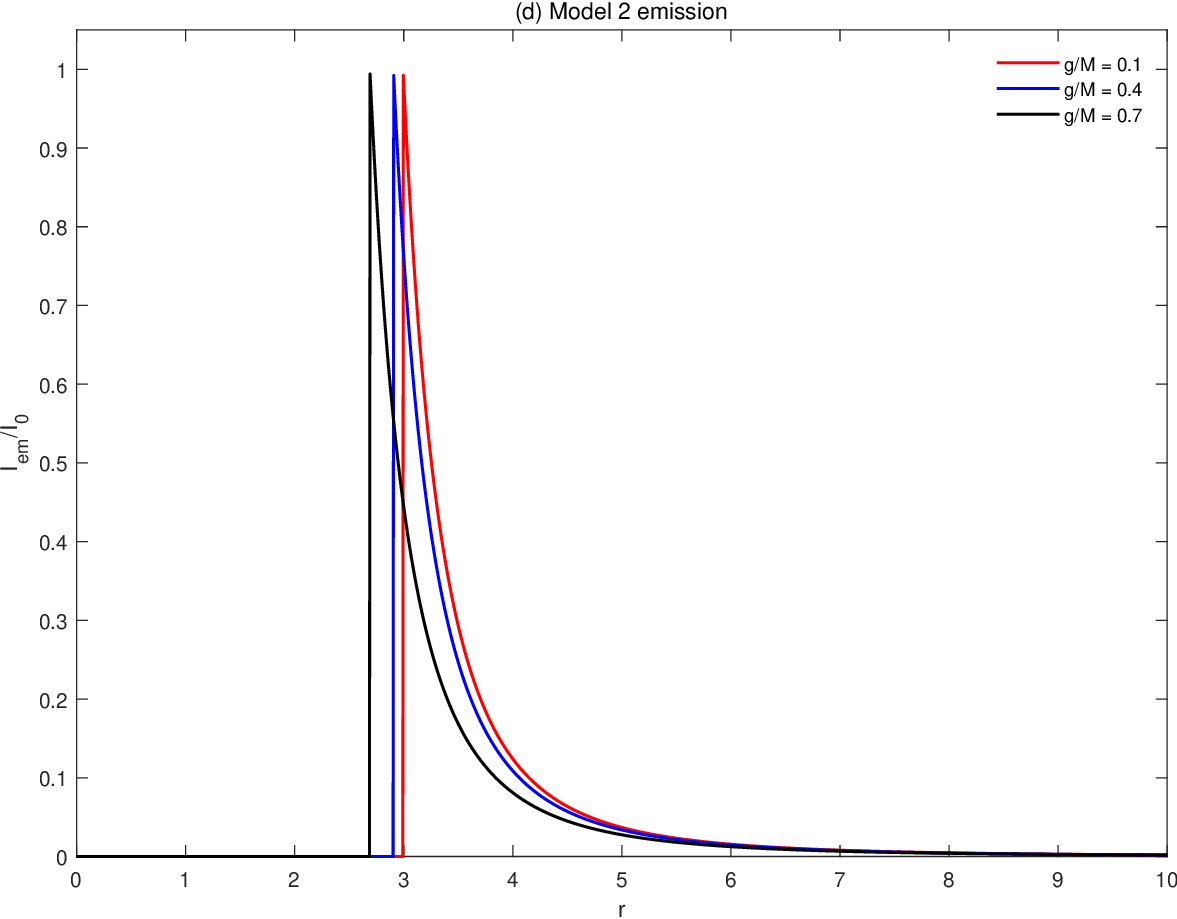}
	\end{minipage}
    \hspace{0.5cm}
    \begin{minipage}[c]{0.3\textwidth}
        \centering
		\includegraphics[scale=0.26,angle=0]{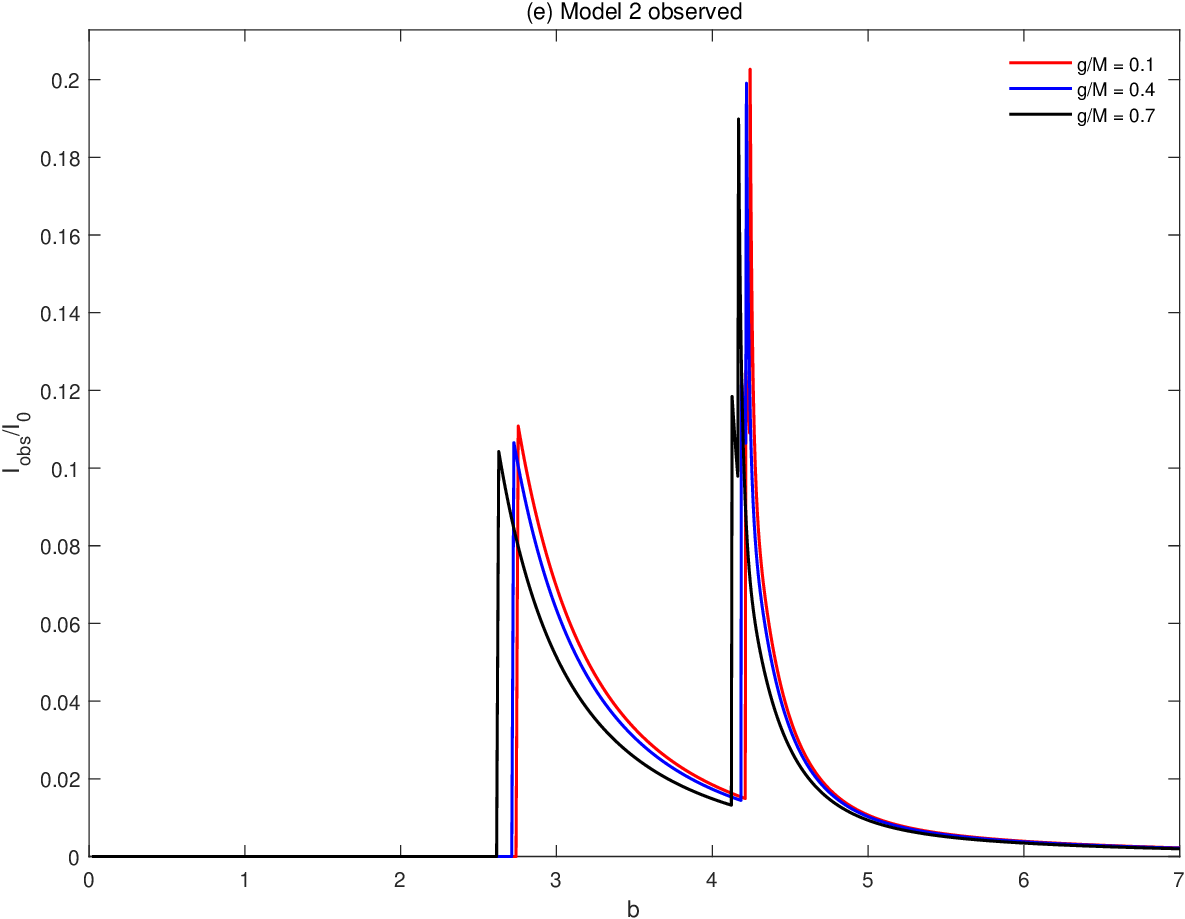}
	\end{minipage}	
   \begin{minipage}[c]{0.34\textwidth}
    \centering
    \includegraphics[height=3.9cm,scale=0.8]{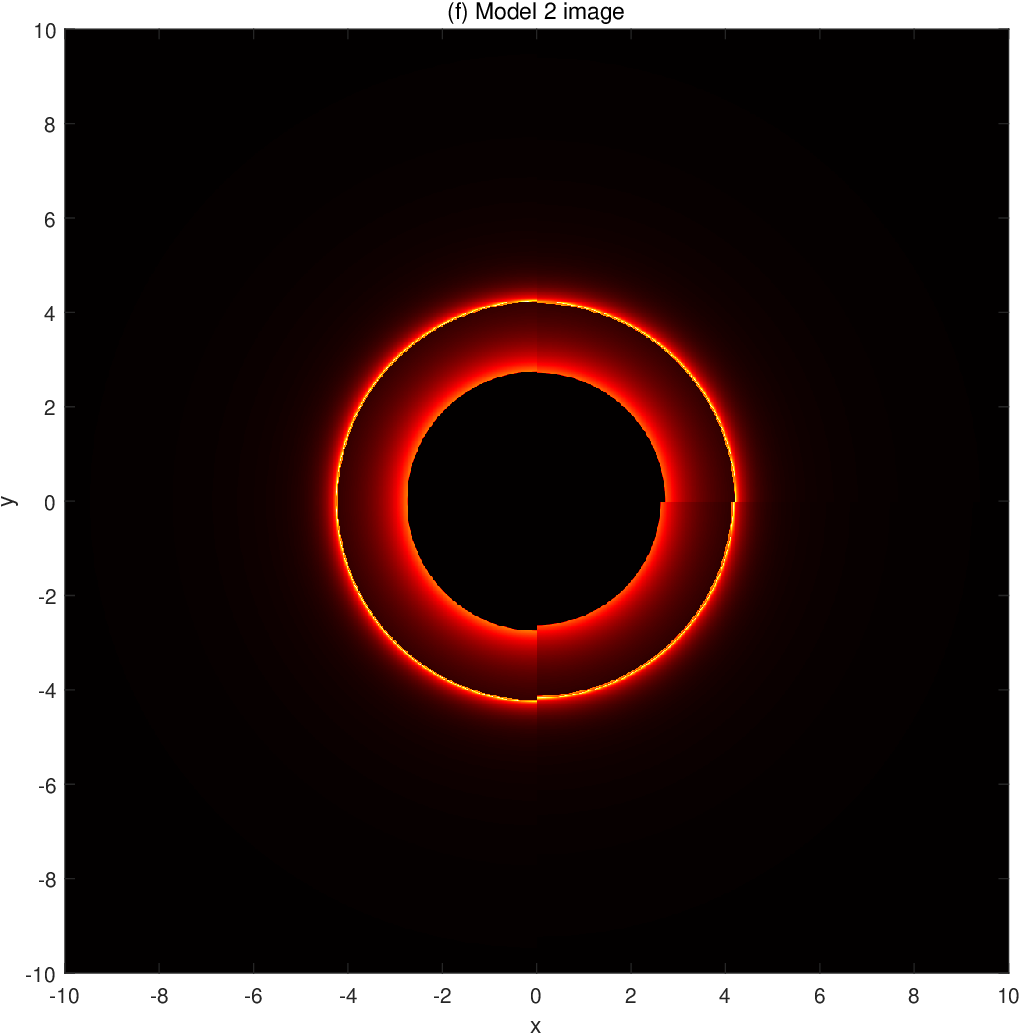}
    \hspace{-0.25cm}
    \raisebox{0cm}{%
        \includegraphics[height=3.9cm,width=0.11\linewidth]{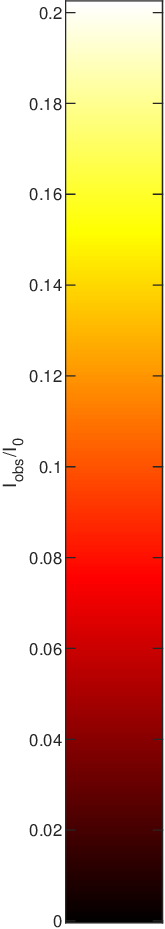}
    }
\end{minipage}

\vspace{2em}

  \centering
	\begin{minipage}[c]{0.3\textwidth}
	    \centering
		\includegraphics[scale=0.26,angle=0]{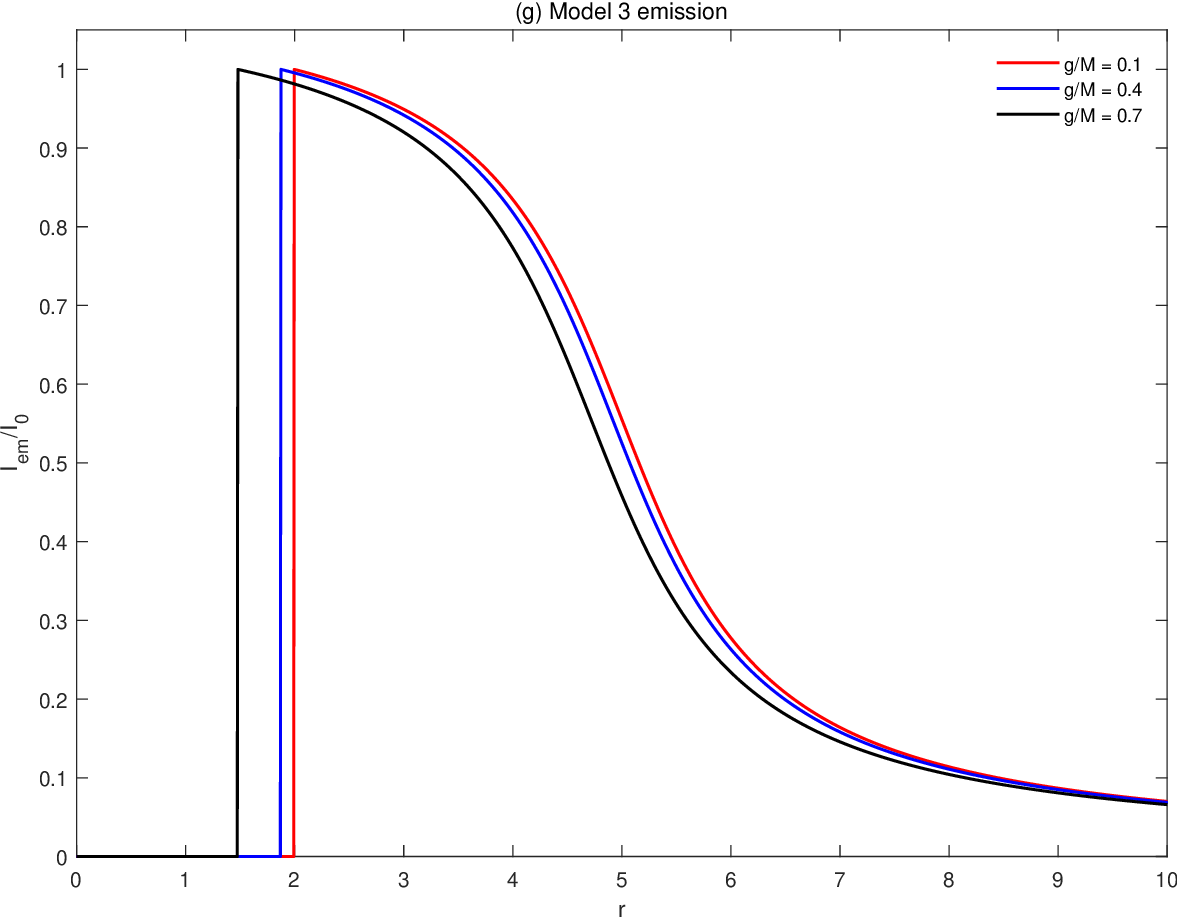}
	\end{minipage}
    \hspace{0.5cm}
    \begin{minipage}[c]{0.3\textwidth}
        \centering
		\includegraphics[scale=0.26,angle=0]{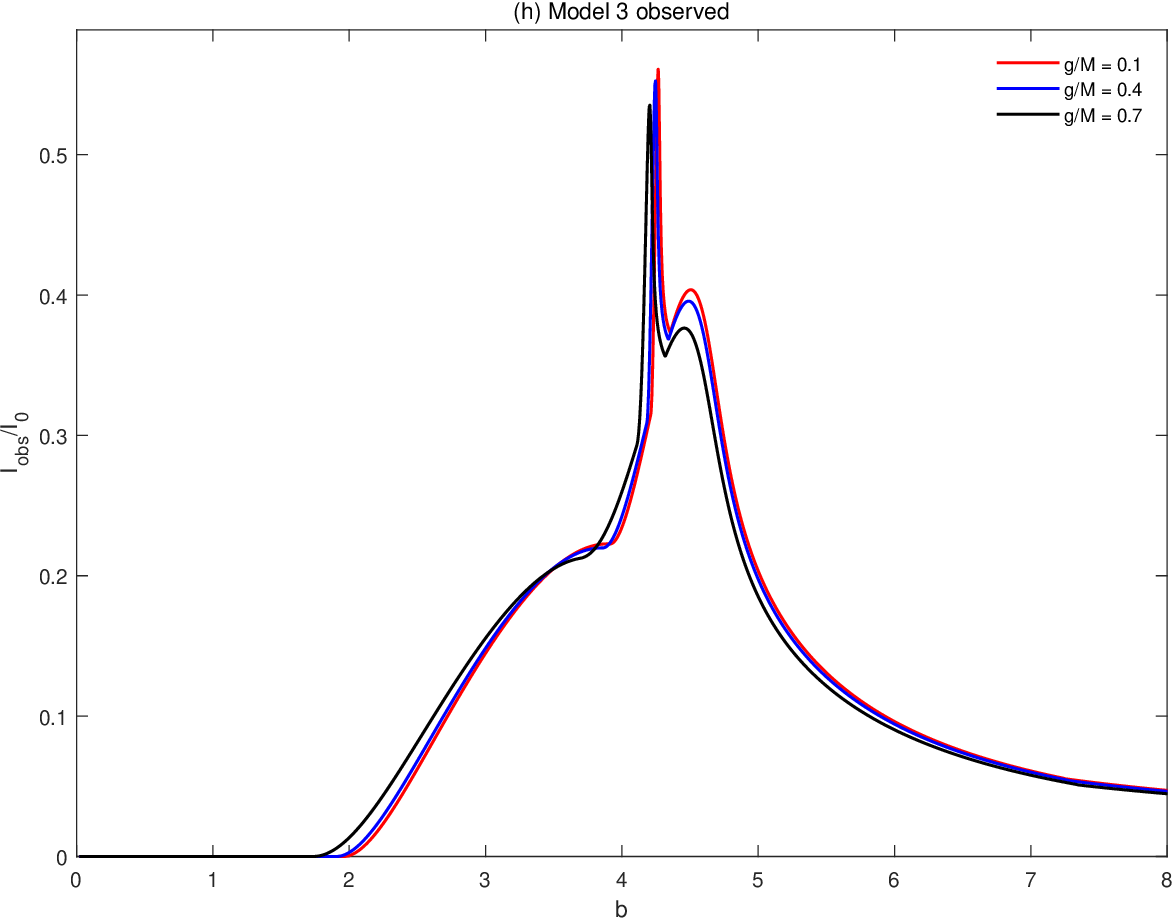}
	\end{minipage}	
   \begin{minipage}[c]{0.34\textwidth}
    \centering
    \includegraphics[height=3.9cm,scale=0.8]{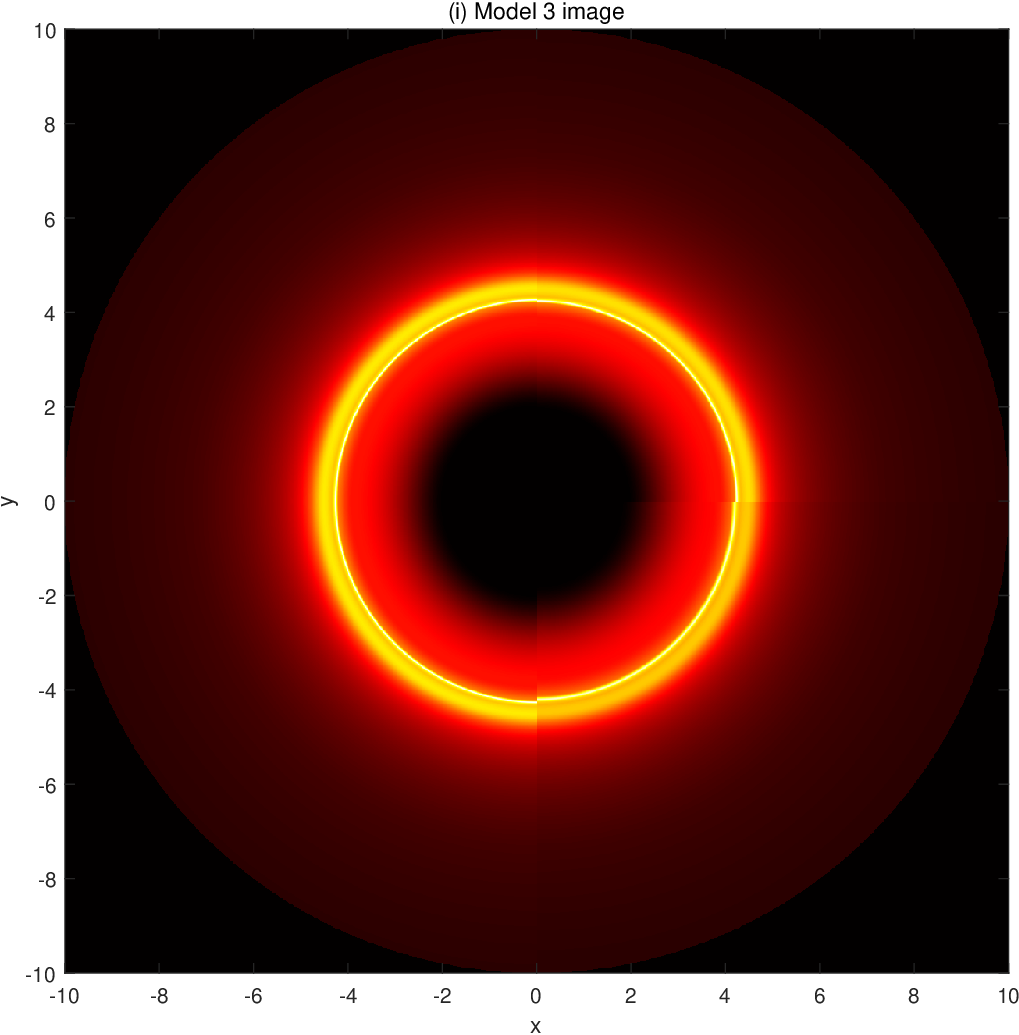}
    \hspace{-0.25cm}
    \raisebox{0cm}{%
        \includegraphics[height=3.9cm,width=0.11\linewidth]{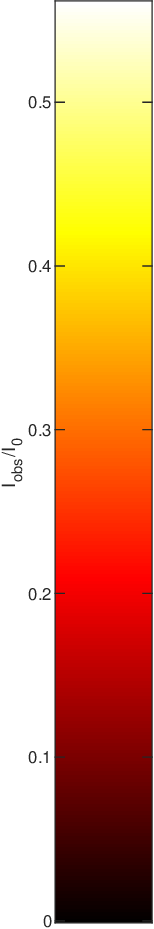}
    }
\end{minipage}
\caption{The face-on optical appearance of the thin accretion disk is shown for three different emission prescriptions. The top, middle, and bottom rows correspond to Models 1, 2, and 3. In each row, the red, blue, and black curves denote configurations with fixed magnetic charges $g/M=0.1$, $0.4$, and $0.7$. In all panels, both the emitted intensity $I_{\rm em}$ and the observed intensity $I_{\rm obs}$ are normalized by the maximum emission intensity $I_{0}$ outside the event horizon, so that the vertical axes are dimensionless.
} \label{fig.6}
\end{figure}

Figure \ref{fig.6} presents the observed emission profiles, intensity distributions, and image-plane mappings for the NLED-Bardeen BH under three representative magnetic charge configurations, $g/M=0.1$, $0.4$, and $0.7$. For each emissivity model, the first column shows the intrinsic radial emissivity profile of the thin accretion disk, while the second column displays the corresponding observed intensity measured by a distant observer. The third column further translates these intensity profiles into two-dimensional images on the celestial coordinate plane, providing a direct visualization of the optical appearance of the BH and its surrounding disk.
In all panels, the red, blue, and black curves correspond to $g/M=0.1$, $0.4$, and $0.7$, respectively. This arrangement allows a systematic comparison of how the magnetic charge affects both the radial brightness distribution and the apparent image morphology. In particular, variations in $g/M$ modify the effective optical geometry experienced by photons, thereby changing the mapping between the emitting disk and the observer's screen. As a result, the peak positions, relative brightness, and ring-like structures in the observed images exhibit clear magnetic charge dependent behavior. The figure therefore provides a useful visual diagnostic for identifying the role of the NLED correction in shaping the observable appearance of the accretion disk, as discussed below.


In the first column, the emitted intensity is controlled by the inner boundary adopted in each model: Model 1 peaks near the $r_{\rm isco}$, Model 2 peaks near the $r_{\rm ph}$, and Model 3 reaches its maximum close to the $r_{\rm +}$. As $g/M$ increases, all characteristic radii, including $r_{\rm +}$, $r_{\rm ph}$, $b_{\rm c}$, and $r_{\rm isco}$, decrease monotonically. This indicates that the magnetic charge compresses the effective gravitational and optical structure of the NLED-Bardeen BH. The second column demonstrates that the observed intensity. For Model 1, the observed intensity profile exhibits three distinct peaks, which can be attributed to the direct emission,lensing ring and photon ring contributions. For $g/M=0.1$, $0.4$, and $0.7$, the maximum peak values of $I_{\rm obs}/I_{0}$ are approximately $0.4871$, $0.4927$, and $0.5218$. Increasing the magnetic charge enhances the observed brightness. This behavior is a direct consequence of the NLED correction, which modifies the effective optical geometry experienced by photons and changes the mapping between the emitting disk and the observer's screen. In Model 2, the observed intensity profile contains two prominent peaks, mainly arising from the direct emission and lensing ring components. The photon ring contribution is comparatively weak and is therefore not clearly visible in the intensity profile. For $g/M=0.1$, $0.4$, and $0.7$, the corresponding maximum peak values are $0.2027$, $0.1996$, and $0.1898$. We can see that Model 1 exhibits more pronounced observational signatures than Model 2. Increasing the magnetic charge leads to a decrease in the observed brightness, which constitutes a characteristic feature distinguishing Model 2 from Model 1. In addition, Model 3 displays a behavior distinct from that of Models 1 and 2. In this model, the disk emission is allowed to extend inward to the event horizon, so the observed intensity includes contributions from the near-horizon strong-field region. However, after removing the overall conformal factor \(1/L_F\), the dominant observed peak of Model 3 does not move to a larger impact parameter. Instead, the peak position shifts slightly inward as the magnetic charge increases: \(b_{\rm peak}=4.2660, 4.2470,\) and \(4.2056\) for \(g/M=0.1, 0.4,\) and \(0.7\). This trend is consistent with the monotonic decrease of the characteristic radii \(r_h\), \(r_{\rm ph}\), \(b_{\rm ph}\), and \(r_{\rm isco}\). Physically, increasing \(g/M\) modifies the optical geometry and reduces the characteristic length scales that control photon propagation. As a result, the emission region and its corresponding image are mapped to slightly smaller impact parameters on the observer’s screen. Meanwhile, the peak intensity of Model 3 also decreases mildly with increasing magnetic charge, from \(0.5608\) to \(0.5352\). Therefore, Model 3 is still more sensitive to the inner disk boundary and strong-field lensing than Models 1 and 2, but its observed profile is now governed by a smoother inward shift and a moderate suppression of the peak intensity as \(g/M\) increases.

The third column translates these radial intensity profiles into two-dimensional face-on images. The bright rings in Models 1 and 2 become more compact as $g/M$ increases, tracking  the inward shift of the dominant observed peak. Meanwhile, Model 3 displays a broader and more extended bright structure because its emission originates from deeper regions close to the horizon. Overall, the magnetic charge acts as a regulator of both the geometric scale and the radiative distribution: it contracts the characteristic optical radii, changes the relative weight of direct and higher-order images, and leaves observable signatures in the peak position, ring compactness, and brightness hierarchy of the accretion disk image. Should this characteristic feature be detected in future astronomical observations, it could provide a means of distinguishing Model 3 from Models 1 and 2 and offer observational support for the validity of Model 3.

\section{Observed optical appearance of NLED–Bardeen black hole with spherical accretions}

Having investigated the optical signatures produced by a geometrically thin accretion disk, we now turn to the case of spherical accretion around the NLED-Bardeen BH. In contrast to the disk configuration, spherical accretion provides a more symmetric emission environment and therefore allows the influence of the NLED-modified photon geometry on the observed image to be identified more directly. In this scenario, radiation emitted throughout the accreting medium is transported along null geodesics of the effective optical metric and is subsequently mapped onto the screen of a distant observer. The resulting intensity distribution is determined by the combined effects of gravitational redshift, photon capture, and the enhanced path length of light rays propagating near the unstable photon sphere. By analyzing the radial intensity profiles and the corresponding two-dimensional images for different values of the magnetic charge $g/M$, we aim to clarify how NLED modifies the apparent shadow size, the location and brightness of the photon ring, and the overall optical morphology relative to the Bardeen and Sch BHs.

\subsection{Static spherical accretion flow}

We first consider an idealized model of spherical accretion, in which the emitting matter is assumed to be static and distributed spherically around the NLED-Bardeen BH. Although this simplified configuration does not capture the full dynamics of a realistic accretion flow, it provides a useful framework for isolating the effects of the effective optical geometry on the observed image. For a distant observer, the specific intensity is obtained by integrating the radiative transfer equation along the photon trajectory $\gamma$, which follows
\be
\label{45}
I_{\mathrm{obs}}=\int_{\gamma} \bar{g}^3 j\left(\nu_{\mathrm{e}}\right) d l_{\mathrm{prop}},
\ee
where $\bar{g}\equiv\nu_{\rm o}/\nu_{\rm e}$ denotes the redshift factor, where $\nu_{\rm e}$ and $\nu_{\rm o}$ are the photon frequencies measured in the emitter frame and by the distant observer, respectively. The quantity $j(\nu_{\rm e})$ represents the emissivity per unit proper volume and per unit frequency evaluated in the local rest frame of the emitting matter, while $dl_{\rm prop}$ is the infinitesimal proper length measured along the photon path. The resulting brightness distribution therefore incorporates both the local emissivity of the accreting medium and the gravitational redshift and lensing effects accumulated along the light ray \cite{Bambi:2013nla,Jaroszynski:1997bw}.

In the NLED-Bardeen BH, the gravitational redshift factor for radiation emitted by static accreting matter and detected by a distant observer is given by $\bar{g}=\sqrt{A(r)}$, where the temporal metric function is normalized at the observer’s location. Assuming monochromatic emission at a characteristic frequency $\nu_{\rm s}$, the specific emissivity can be modeled in the form given by 
\be
\label{46}
j\left(\nu_{\mathrm{e}}\right) \propto \delta\left(\nu_e-\nu_s\right) r^{-m},
\ee
here we adopt the emissivity index $m=2$, following the conventional choice in Ref. \cite{Bambi:2013nla}. Other commonly used values, including $m=4$ \cite{Kocherlakota:2022jnz} and $m=6$ \cite{Bauer:2021atk}, have likewise been shown to provide physically reasonable descriptions of the emitting medium. The infinitesimal proper length can then be written as
\be
\label{47}
dl_{\rm prop}=\sqrt{B(r)+\frac{C(r)}{\left(\frac{dr}{d\phi}\right)^2}} dr,
\ee
where the function $dr/d\phi$ is given by Eq.~\eqref{19}. Consequently, the specific intensity measured by a static observer can be expressed as
\be
\label{48}
I_{\rm obs}(b)=\int_\gamma\frac{A(r)^{3/2}}{r^2}\sqrt{B(r)
+\frac{1}{C(r)\left[\frac{1}{A(r)B(r)b^2}-\frac{1}{B(r)C(r)}\right]}} dr.
\ee

Equation \eqref{48} expresses the total observed intensity as a function of the impact parameter $b$. Since the impact parameter determines the photon trajectory and hence the propagation path through the emitting medium, variations in $b$ directly modify the specific intensity measured by a distant observer. We therefore investigate the dependence of the observed intensity on $b$ in order to clarify how different photon trajectories contribute to the optical appearance of the BH.

\begin{figure}[htbp]
    \centering

    \makebox[\textwidth][c]{%
        \begin{minipage}[c]{0.45\textwidth}
            \centering
            \includegraphics[
                width=\linewidth
            ]{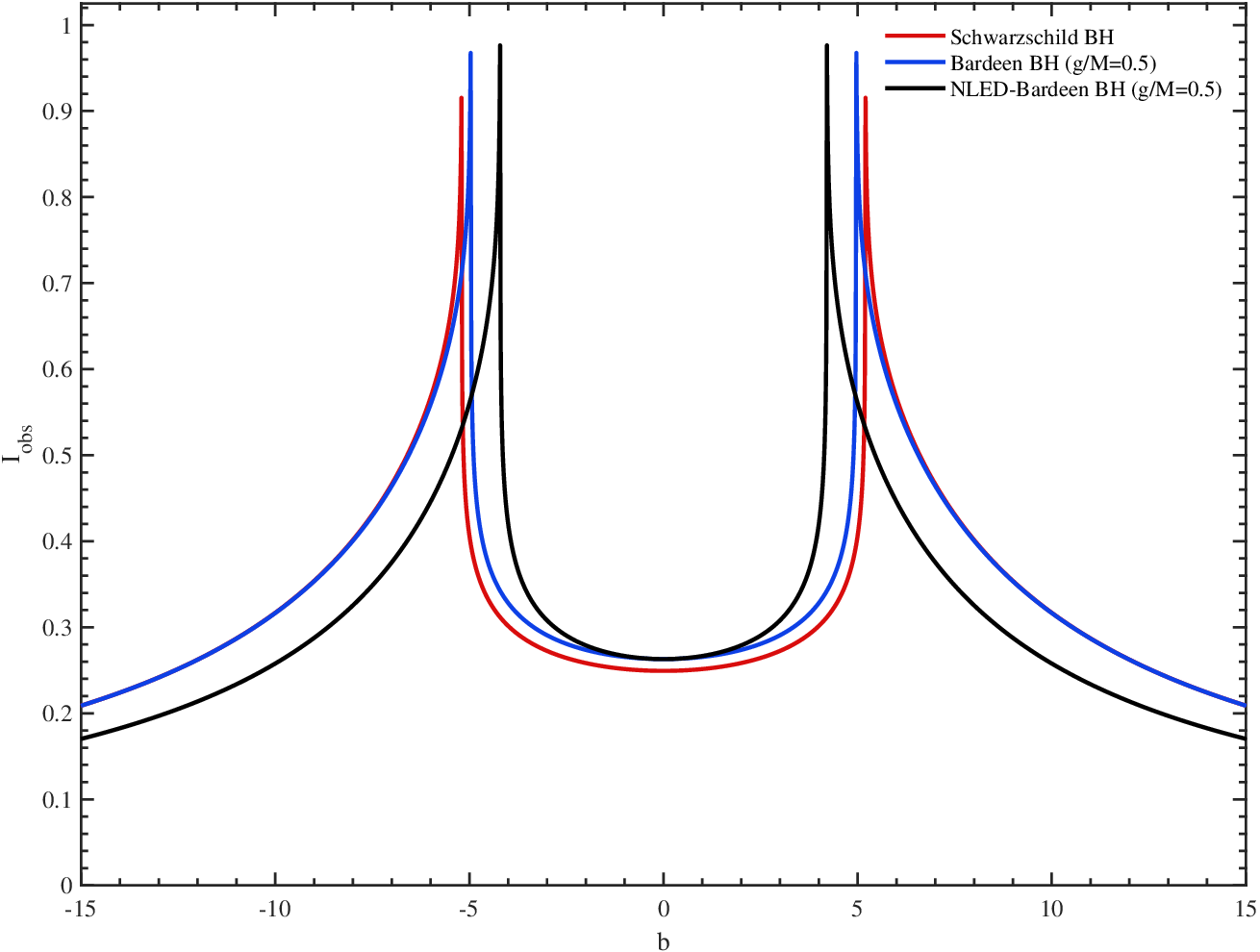}
        \end{minipage}
        \hspace{0.07\textwidth}
        %
        \begin{minipage}[c]{0.49\textwidth}
            \centering
            \includegraphics[
                height=5.8cm
            ]{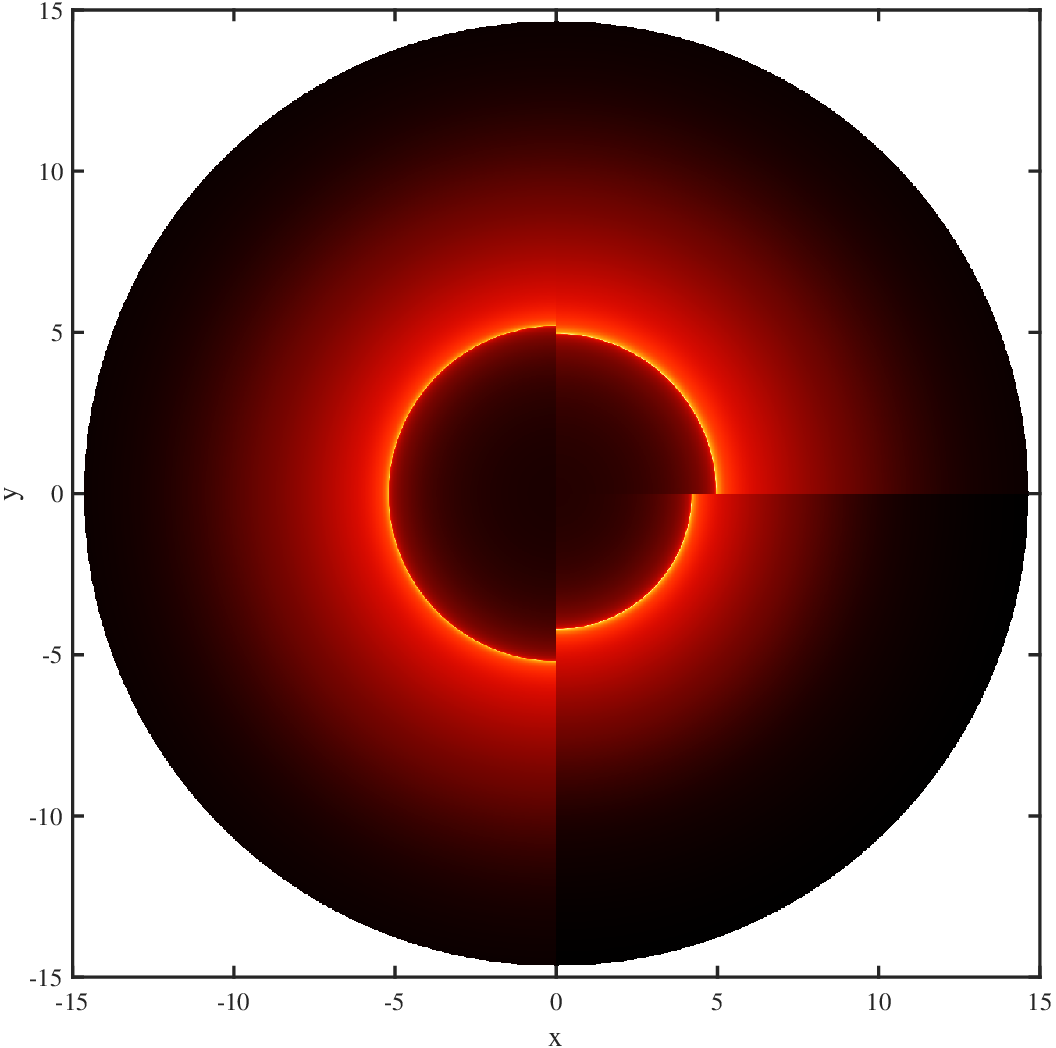}
            \hspace{0.15cm}
            \includegraphics[
                height=5.8cm,
                width=0.055\linewidth
            ]{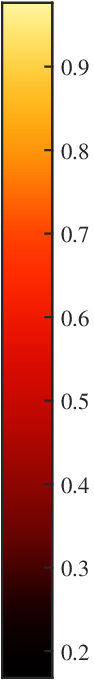}
        \end{minipage}%
    }

    \vspace{1.5em}

    \makebox[\textwidth][c]{%
        \begin{minipage}[c]{0.45\textwidth}
            \centering
            \includegraphics[
                width=\linewidth
            ]{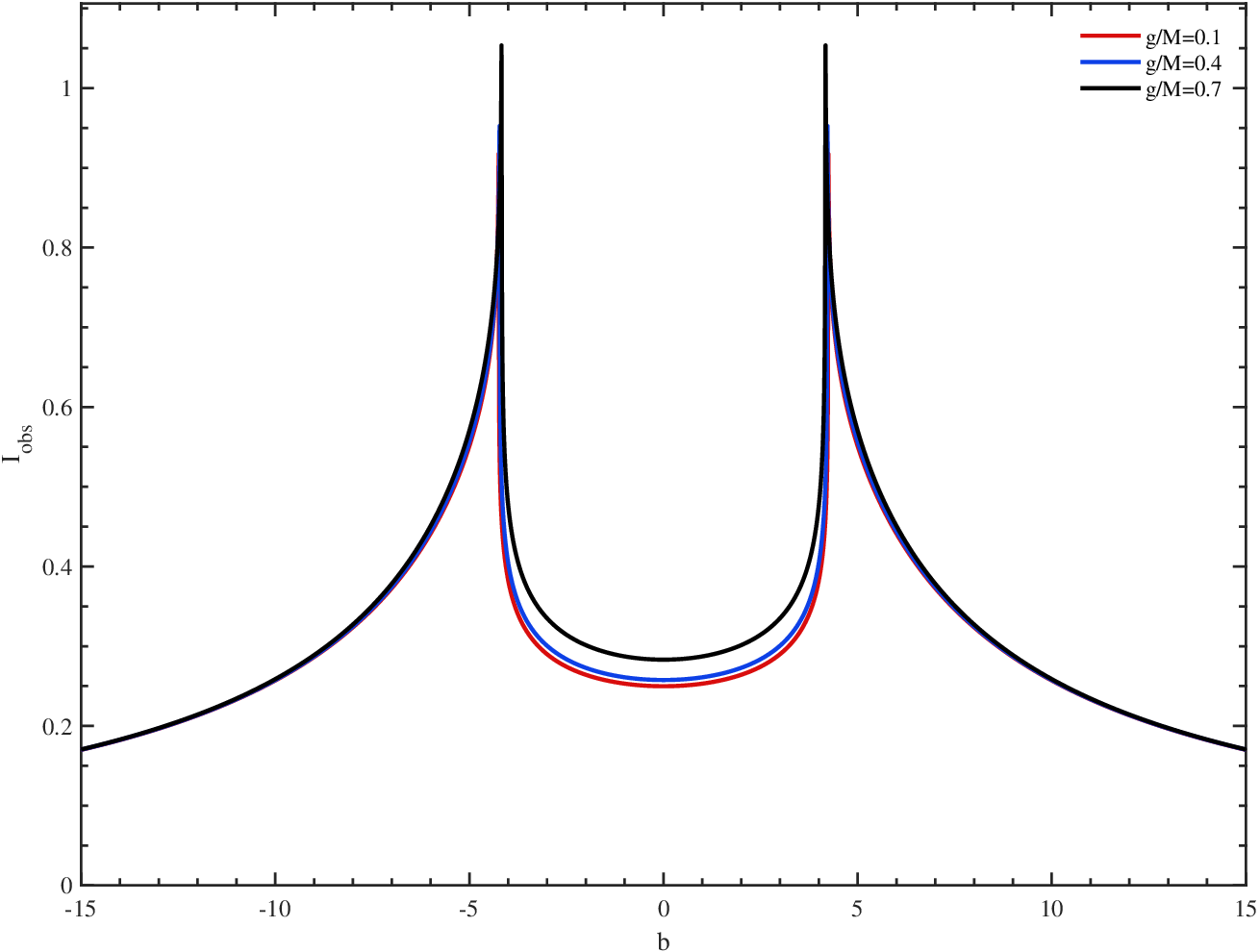}
        \end{minipage}
        \hspace{0.07\textwidth}
        %
        \begin{minipage}[c]{0.49\textwidth}
            \centering
            \includegraphics[
                height=5.8cm
            ]{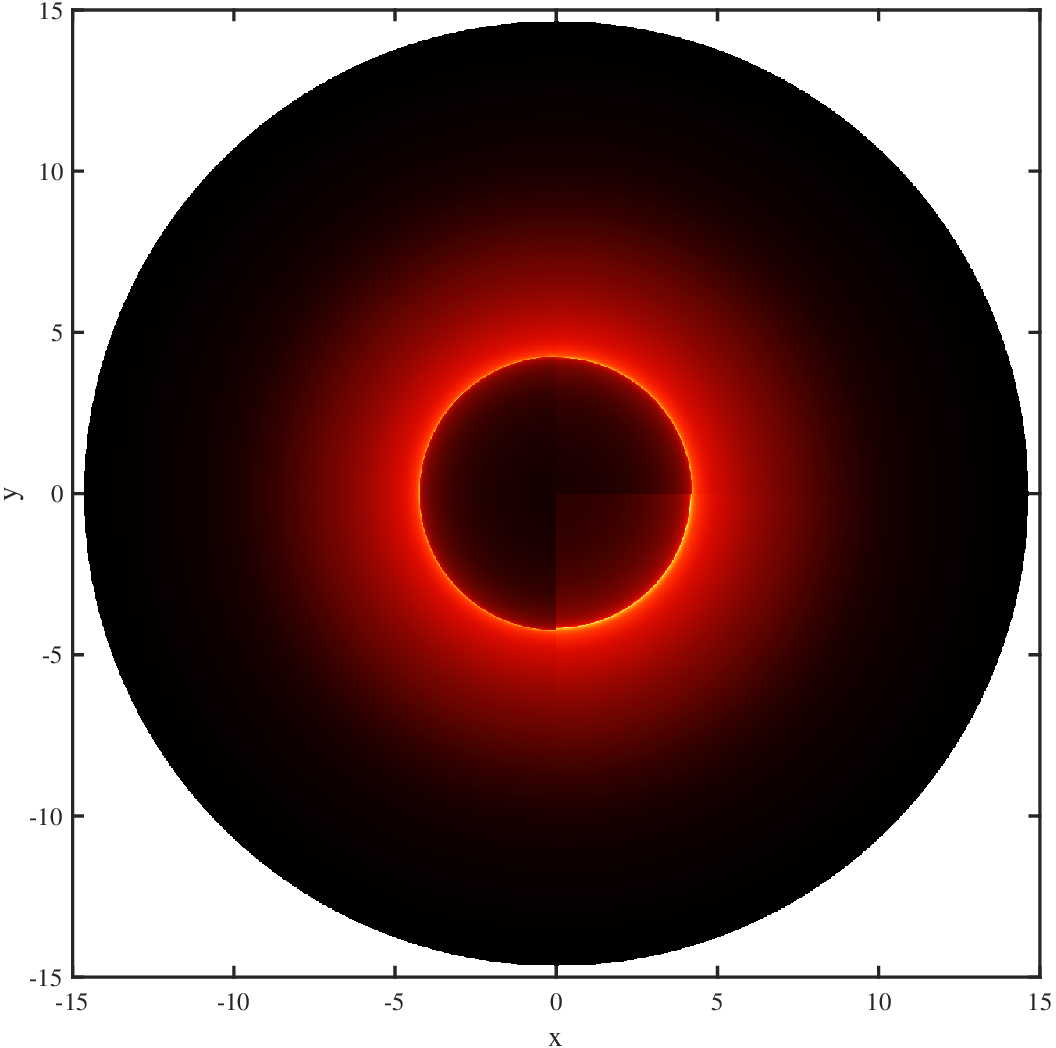}
            \hspace{0.15cm}
            \includegraphics[
                height=5.8cm,
                width=0.055\linewidth
            ]{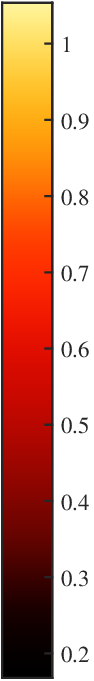}
        \end{minipage}%
    }

    \caption{
    Comparison of the observed intensity profiles and the corresponding two-dimensional images for static spherical accretion. The upper panels compare the Sch, Bardeen, and NLED-Bardeen BHs, whereas
    the lower panels display the NLED-Bardeen BH for different values of the magnetic charge.}
    \label{fig.7}
\end{figure}

Figure~\ref{fig.7} compares the observed intensity profiles $I_{\rm obs}$ and the corresponding images produced by spherical accretion around different BHs. The upper panels present the results for the Sch BH, the Bardeen BH with $g/M=0.5$, and the NLED-Bardeen BH with $g/M=0.5$, while the lower panels compare the NLED-Bardeen configurations with $g/M=0.1$, $0.4$, and $0.7$. For $b<b_{\rm c}$, the observed intensity initially increases gradually with the impact parameter $b$ and then rises sharply as $b$ approaches the critical value $b_{\rm c}$. The intensity reaches its maximum in the $b=b_{\rm c}$, where photons propagate close to the unstable photon sphere and maximizing the observed intensity.
For $b>b_{\rm c}$, the observed intensity decreases monotonically with increasing $b$ and asymptotically approaches zero as $b\rightarrow\infty$. This characteristic peak near $b_{\rm c}$ gives rise to the bright ring surrounding the central dark region in the corresponding two-dimensional images.


Under the same spherical accretion flow model, the peak values of $I_{\rm obs}$ for the Sch, Bardeen with $g/M=0.5$, and NLED-Bardeen BHs are approximately $0.915$, $0.968$, and $0.976$, respectively. This ordering shows that the magnetic charge enhances the observed emission and that the additional NLED correction produces a further, although moderate, increase in the peak intensity. Accordingly, the NLED correction yields the highest observed intensity for the NLED-Bardeen BH, followed by the Bardeen BH and, finally, the Sch BH. A similar trend is found within the NLED-Bardeen family: as $g/M$ increases from $0.1$ to $0.4$ and $0.7$, the peak intensity rises from approximately $0.917$ to $0.952$ and $1.054$, in agreement with the progressively brighter and more concentrated rings displayed in the corresponding two-dimensional images. This behavior can be attributed to the inward shift of the photon sphere with increasing magnetic charge. Near-critical photons then propagate over a longer effective optical path through the emitting medium, and the resulting enhancement of the accumulated radiation gradually outweighs the attenuation caused by gravitational redshift, thereby producing a higher and more sharply localized intensity peak.

\subsection{Infalling spherical accretion flow}

In this subsection, we extend the analysis to an radially infalling accretion flow. Compared with the static spherical configuration, this dynamical model provides a more realistic description of astrophysical accretion, since the surrounding matter is generally expected to possess a nonvanishing inward velocity. The formal expression for the observed specific intensity given in Eq. \eqref{45} remains applicable; however, the redshift factor must now be modified to account for the motion of the emitting matter. In this scenario, the redshift factor therefore depends not only on the gravitational field but also on the four-velocity of the accretion flow, which follows \cite{Wang:2023vcv,Bambi:2013nla,KumarWalia:2024yxn,Luo:2025xjb}
\be
\label{49}
\bar{g}=\frac{p_\rho u_o^\rho}{p_\sigma u_e^\sigma},
\ee
where $p_{\rho}$ denotes the four-momentum of photons emitted by the accreting matter. The four-velocity of a distant observer is given by $u_{\rm o}^{\mu}=\left(1,0,0,0\right)$. For matter undergoing radial free fall from rest at infinity, the four-velocity of the emitting accretion flow can be written as $u_{\rm e}^{t}=1/A(r)$, $u_{\rm e}^{r}=-\sqrt{\frac{1-A(r)}{A(r)B(r)}}$, and $u_{\rm e}^{\theta}=u_{\rm e}^{\phi}=0$, where the negative sign of $u_{\rm e}^{r}$ indicates inward radial motion toward the NLED-Bardeen BH.

Upon rescaling the affine parameter according to $\lambda\rightarrow \lambda/|L|$, the conserved temporal component of the wave vector can be normalized as $p_t=1/b$. The corresponding radial component $p_r$ is then obtained by imposing the null condition $p_{\alpha}p^{\alpha}=0$, which yields
\be
\label{50}
\frac{p_r}{p_t}= \pm \sqrt{\frac{B(r)}{A(r)}-\frac{b^2B(r)}{C(r)}},
\ee
here the $\pm$ correspond to photons propagating radially outward and inward relative to the NLED-Bardeen BH. For brevity, we display only the two components of the photon four-momentum required in the calculation. Substituting the expressions in Eq. \eqref{50} into Eq. \eqref{49}, the redshift factor for the radially infalling accretion flow can be obtained as
\be
\label{51}
\bar{g}=\frac{1}{u_{\rm e}^{t}+(p_r/p_t)u_{\rm e}^{r}},
\ee
which differs fundamentally from the static case because of the radial infall of the accreting fluid. Furthermore, the infinitesimal proper length element can be defined as \cite{Wang:2023vcv,Bambi:2013nla}
\be
\label{52}
d l_{\mathrm{prop}}=p_\mu u_{\mathrm{e}}^\mu d \lambda=\frac{p_t}{\bar{g}\left|p_r\right|} dr.
\ee

For the emissivity, we retain the monochromatic emission approximation, so that Eq. \eqref{46} remains applicable. By integrating Eq. \eqref{45} over the observed frequency, the observed flux for the distant  observer  in  the  case  of  infalling  spherical accretion model is 
\be
\label{53}
I_{\rm obs}(b)=\int_\gamma \frac{\bar{g}^3 p_t}{r^2\left|p_r\right|} dr.
\ee

Equation \eqref{53} shows that the observed intensity in the infalling spherical accretion model is obtained by integrating the radiative contribution along the complete null geodesic. The factor \(\bar{g}^{\,3}\) accounts for both gravitational redshift and the Doppler effect induced by the radial motion of the emitting matter. 
As shown in Fig. 8, the magnetic charge and the NLED correction primarily shift the intensity peak toward smaller impact parameters, reflecting the reductions in both the photon sphere radius and the critical impact parameter and thereby producing a more compact shadow and photon ring. At the fixed value \(g/M=0.5\), the NLED-Bardeen BH exhibits a noticeably smaller ring radius and a slightly lower peak intensity than the Bardeen BH. Within the NLED-Bardeen family, increasing $g/M$ from $0.1$ to $0.7$ continuously shifts the intensity peak inward, while its amplitude increases monotonically. In comparison with Model 1, the peak intensity is reduced from approximately \(0.92\!-\!1.05\) in the static spherical flow to approximately \(0.45\!-\!0.48\) in the infalling flow. This suppression originates from the additional Doppler redshift experienced by photons emitted by inward-moving matter and escaping toward a distant observer. Since this attenuation enters the transfer equation through the cubic factor \(\bar{g}^{\,3}\), it considerably offsets the enhancement associated with the increased optical path length near the photon sphere, resulting in an overall dimmer appearance for the infalling accretion flow.

\begin{figure}[htbp]
    \centering

    \makebox[\textwidth][c]{%
        \begin{minipage}[c]{0.45\textwidth}
            \centering
            \includegraphics[
                width=\linewidth
            ]{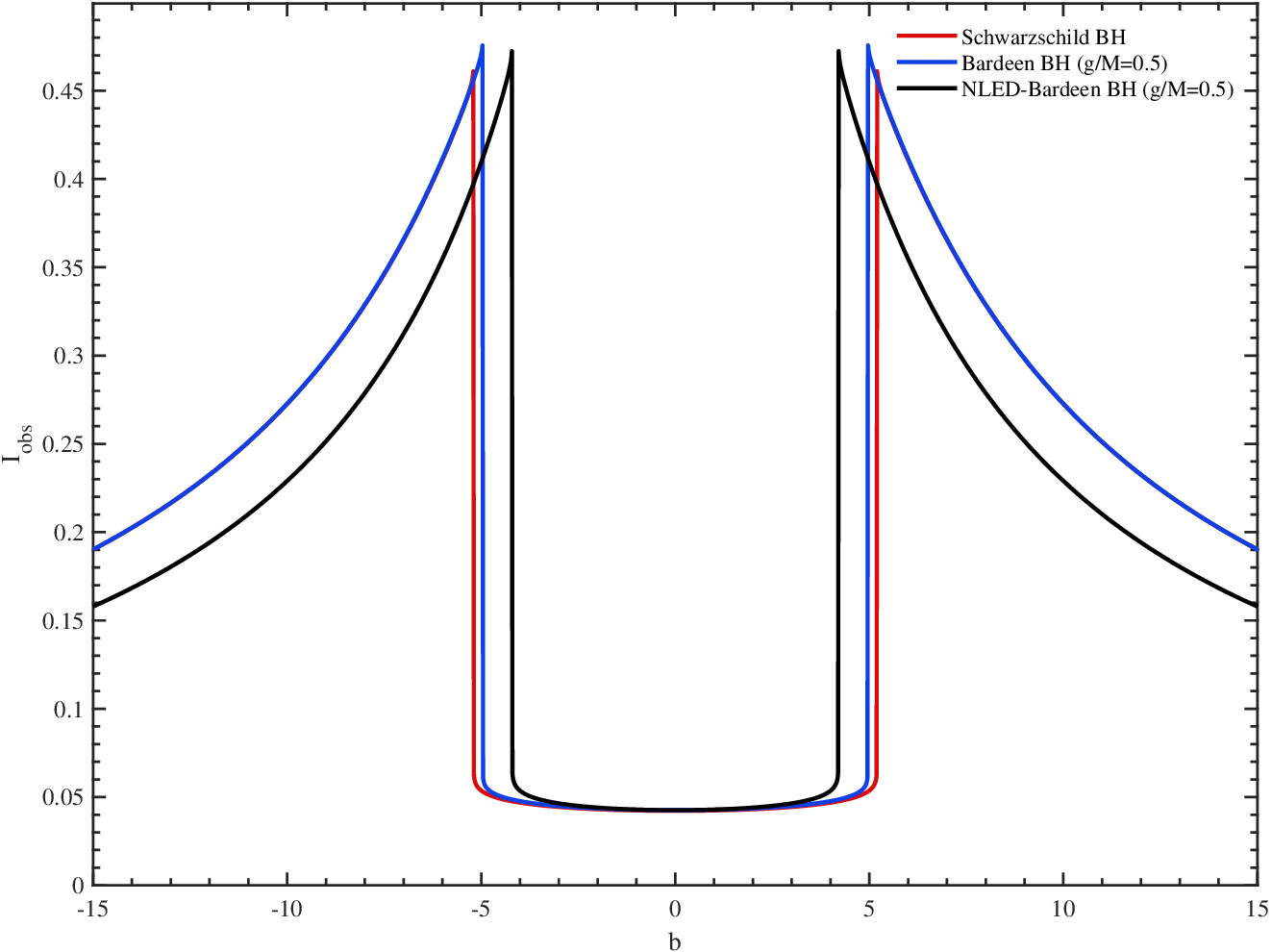}
        \end{minipage}
        \hspace{0.07\textwidth}
        %
        \begin{minipage}[c]{0.49\textwidth}
            \centering
            \includegraphics[
                height=5.8cm
            ]{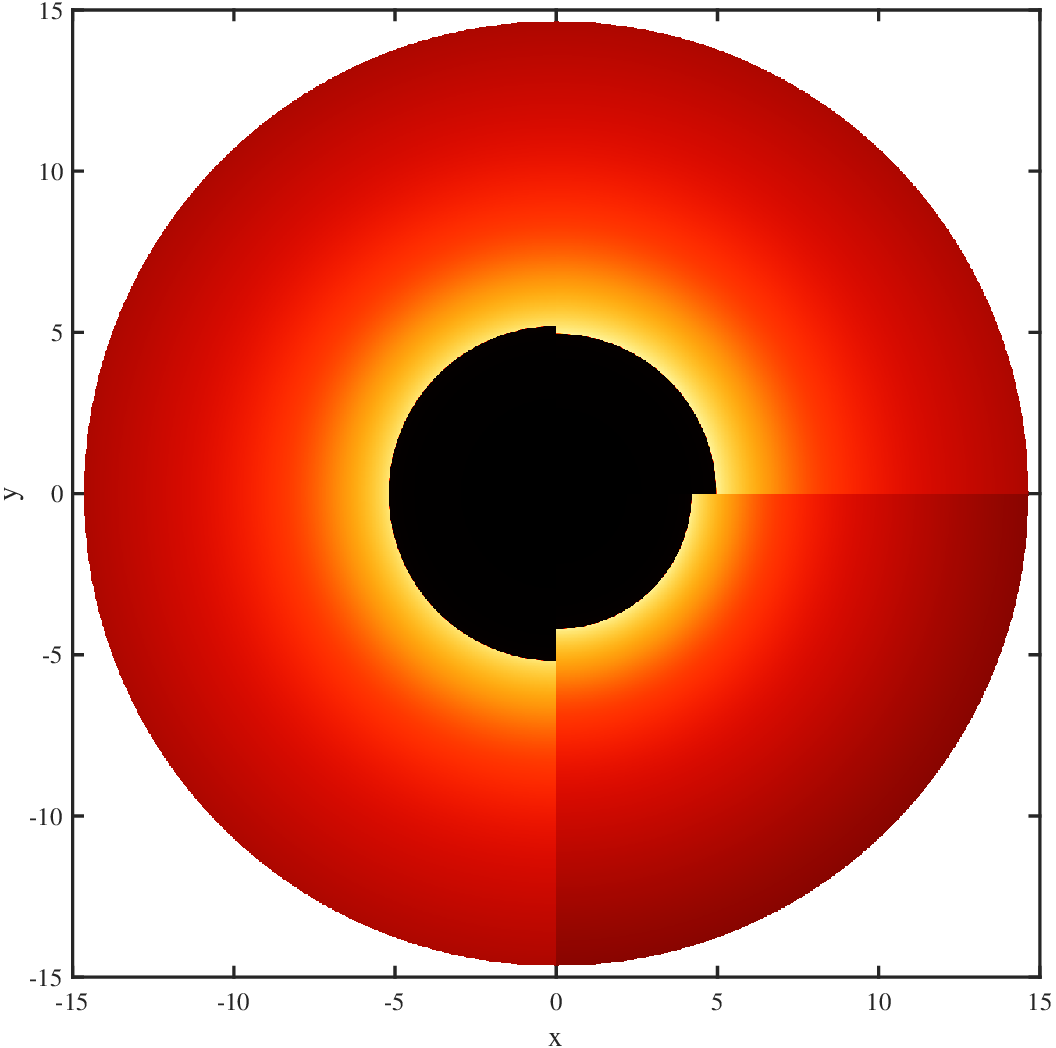}
            \hspace{0.15cm}
            \includegraphics[
                height=5.8cm,
                width=0.055\linewidth
            ]{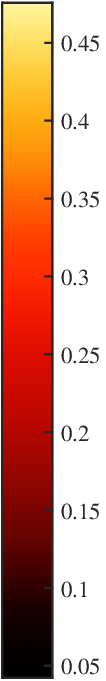}
        \end{minipage}%
    }

    \vspace{1.5em}

    \makebox[\textwidth][c]{%
        \begin{minipage}[c]{0.45\textwidth}
            \centering
            \includegraphics[
                width=\linewidth
            ]{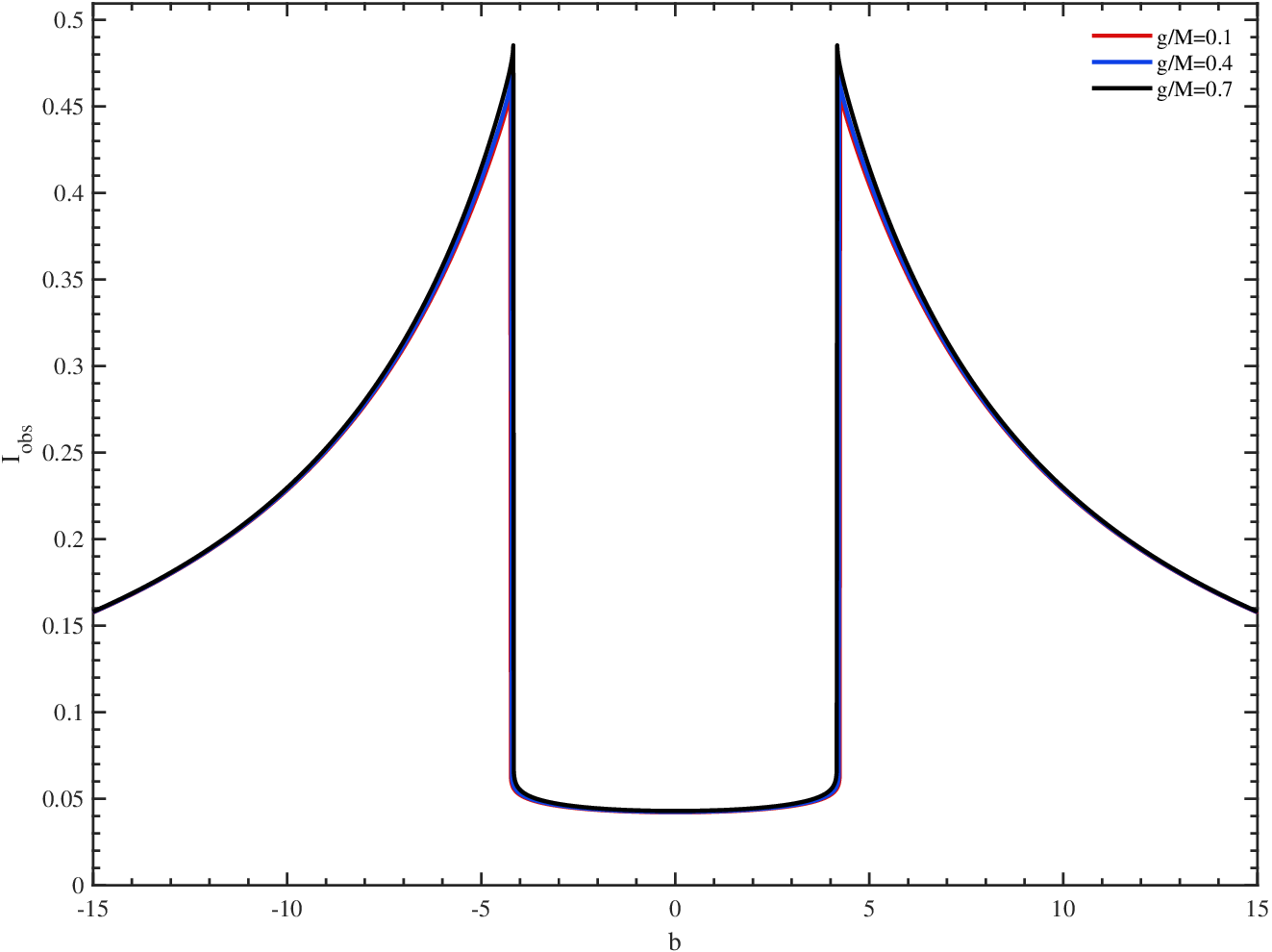}
        \end{minipage}
        \hspace{0.07\textwidth}
        %
        \begin{minipage}[c]{0.49\textwidth}
            \centering
            \includegraphics[
                height=5.8cm
            ]{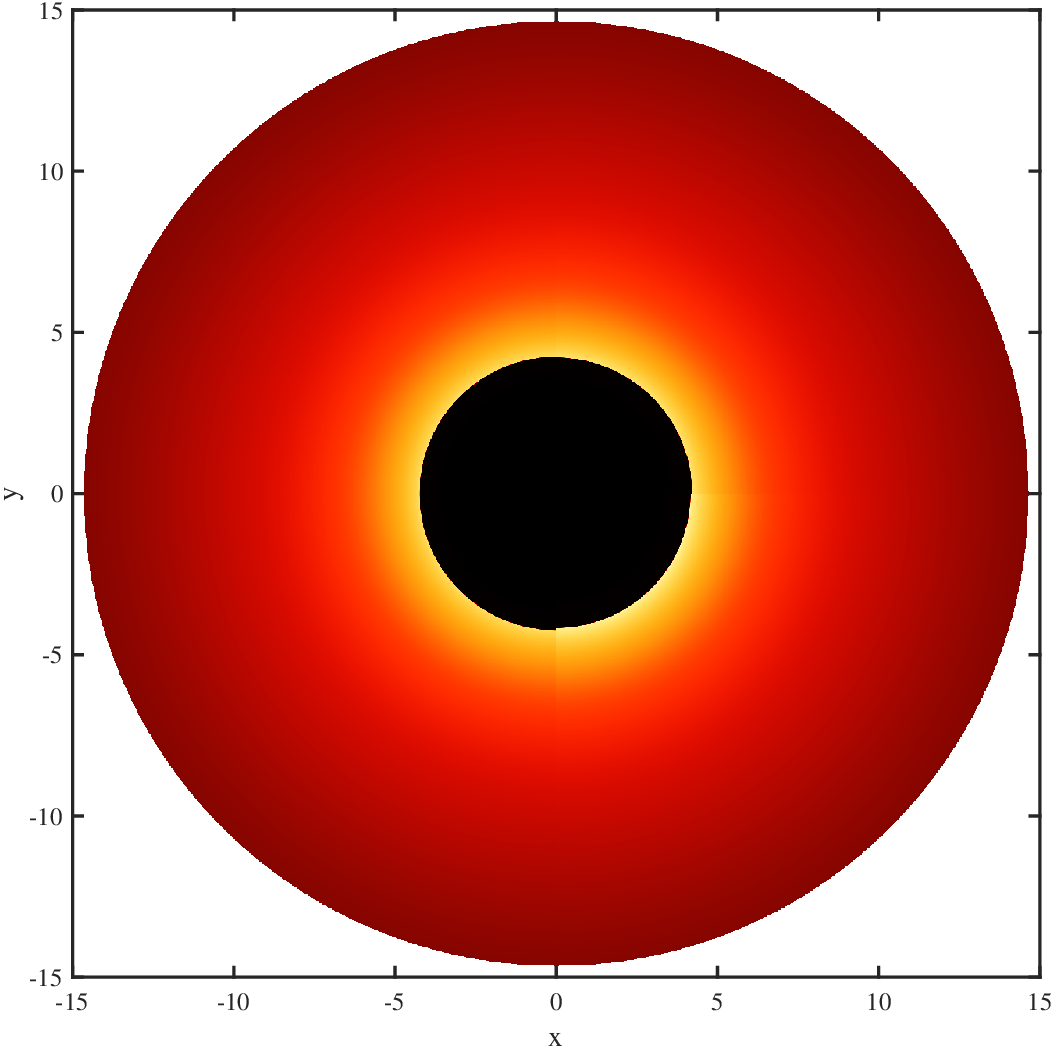}
            \hspace{0.15cm}
            \includegraphics[
                height=5.8cm,
                width=0.055\linewidth
            ]{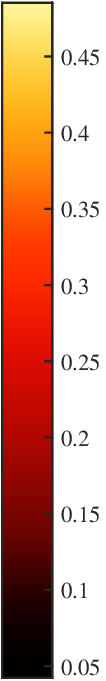}
        \end{minipage}%
    }

    \caption{The left panels show the observed intensity profiles $I_{\rm obs}(b)$ produced by  infalling spherical accretion. We consider the Sch BH, the Bardeen BH with $g/M=0.5$, and the NLED-Bardeen BH with $g/M=0.5$, together with NLED-Bardeen configurations characterized by different magnetic charges. The right panels present the corresponding composite two-dimensional images, allowing the optical appearances of the three BH configurations and the influence of the magnetic charge on the observed brightness distribution to be compared directly.
}
    \label{fig.8}
\end{figure}

\section{Conclusion}

In this work, we investigated the shadow and optical appearance of the NLED-Bardeen BH by consistently describing photon propagation within the framework of nonlinear electromagnetic field. We identified the parameter region that admits a regular event horizon and clarified how the NLED correction, characterized primarily by the dimensionless magnetic charge $g/M$, modifies the optical geometry and, consequently, the observable properties of the BH. 

Our null geodesic analysis shows that the magnetic charge shifts the characteristic strong-field scales inward. As \(g/M\) increases, the event horizon radius, photon sphere radius, critical impact parameter, and observed shadow radius decrease monotonically. Nevertheless, the effects of the background Bardeen geometry and the NLED optical correction must be distinguished carefully. At \(g/M=0.5\), the Bardeen and NLED-Bardeen BHs have the same event horizon, whereas their photon spheres and shadows differ because photons follow different optical metrics. In particular, the photon sphere radii satisfy \(r_{\rm ph}^{\rm Bardeen}<r_{\rm ph}^{\rm NLED-Bardeen}<r_{\rm ph}^{\rm Schwarzschild}\). Although the NLED-Bardeen BH has a smaller critical impact parameter than the Bardeen BH, its effective optical metric does not possess the standard asymptotic normalization. Consequently, the shadow radius measured by a distant observer is \(R_{\rm sh}=\sqrt{3/2}\,b_{\rm c}\), rather than simply \(R_{\rm sh}=b_{\rm c}\). This produces \(R_{\rm sh}/M\simeq5.148\) for the NLED-Bardeen BH at \(g/M=0.5\), which lies between the corresponding Bardeen and Sch values. The same separation between the background and optical geometries also explains why the complete horizon-photon sphere-shadow inequality is not valid over the entire parameter range: the bounds associated with \(X\) and \(Y\) remain satisfied, whereas the horizon-photon sphere relation represented by \(Z\geq1\) can be violated.

The classification of photon trajectories into direct emission, lensing-ring, and photon-ring components further demonstrates how NLED modifies the image formation process. The direct image provides the dominant contribution to the observed intensity, while the higher-order images are strongly suppressed by the steep slopes of their transfer functions. For \(g/M=0.5\), the first-order transfer function of the NLED-Bardeen BH is steeper than those of the Sch and Bardeen BHs, indicating stronger contribution for radial demagnification. Within the NLED-Bardeen family, this slope increases with \(g/M\), so that the direct-image contribution becomes progressively more compressed on the observer’s screen. The resulting brightness variation, however, is not universal and depends sensitively on the assumed emissivity profile. In the thin-disk Model 1, the maximum intensity increases from approximately \(0.4871\) to \(0.5218\) as \(g/M\) increases from \(0.1\) to \(0.7\). By contrast, the maximum intensity decreases from approximately \(0.2027\) to \(0.1898\) in Model 2 and from \(0.5608\) to \(0.5352\) in Model 3. Thus, the inward displacement and increasing compactness of the ring structures are robust geometric effects, whereas the variation of their brightness is controlled jointly by the optical geometry, the disk inner boundary, and the radial emissivity distribution.

The static spherical accretion results provide a complementary physical interpretation. In the static model, the observed intensity reaches a sharp maximum near \(b=b_{\rm c}\), because near-critical photons remain close to the unstable photon sphere and accumulate radiation over an extended optical path. The peak intensities for the Sch, Bardeen, and NLED-Bardeen BHs are approximately \(0.915\), \(0.968\), and \(0.976\), respectively. Within the NLED-Bardeen family, the peak increases from approximately \(0.917\) to \(1.054\) as \(g/M\) increases from \(0.1\) to \(0.7\), indicating that the enhancement of the effective optical path can outweigh gravitational-redshift attenuation in a static emitting medium. Infalling spherical accretion changes this competition substantially. The motion of the emitting matter introduces an additional Doppler redshift through the redshift factor \(\bar g\), whose cubic power enters the radiative transfer integral. The peak intensity is consequently reduced to approximately \(0.45\!-\!0.48\), nearly one-half of that obtained for the static flow. Nevertheless, the peak remains located close to the critical impact parameter and moves inward with increasing magnetic charge, producing a progressively more compact shadow and bright ring. These results demonstrate that the ring position is primarily determined by the effective photon geometry, whereas its brightness is strongly dependent on the velocity field of the accreting matter.

Overall, the most robust signatures of the NLED-Bardeen BH are the modified relation between the photon sphere, critical impact parameter, and observed shadow radius, together with the inward displacement and compression of the lensing structures as the magnetic charge increases. Intensity measurements alone are less model-independent because they depend on the assumed emissivity, the location of the emitting inner boundary, and whether the accreting matter is static or infalling. The present calculations are based on idealized thin disk and spherical accretion flows with prescribed, monochromatic emissivities, while absorption, scattering, plasma dispersion, rotation, inclination, polarization, and self-consistent magnetohydrodynamics are neglected. Therefore, the results should be regarded as theoretical optical diagnostics rather than direct observational constraints. A quantitative comparison with EHT observations will require GR radiative transfer calculations and a statistical analysis that consistently incorporates both spacetime parameters and accretion flow uncertainties.

\begin{acknowledgments}
This study was supported by the Basic Research Program of Shanxi Province (Grant No.202503021211204).
\end{acknowledgments}

\bibliographystyle{unsrt}
\bibliography{EMDAFF}

\end{document}